\documentclass{optica-article}
\journal{boe}
\articletype{Research Article}

\usepackage{amsmath}
\usepackage{bm}
\usepackage{color}
\definecolor{darkred}{rgb}{0.8, 0.0, 0.0}
\definecolor{darkblue}{rgb}{0.2, 0.2, 0.6}
\definecolor{brown(traditional)}{rgb}{0.59, 0.29, 0.0}
\definecolor{auburn}{rgb}{0.43, 0.21, 0.1}
\definecolor{burntorange}{rgb}{0.8, 0.33, 0.0}
\usepackage{xspace}
\usepackage{comment}

\newcommand{\etal}{\textit{et al.\@}\xspace}
\newcommand{\enface}{\textit{en face}\xspace}
\newcommand{\Enface}{\textit{En face}\xspace}
\newcommand{\um}{$\muup \text{m}$\xspace}

\newcommand{\bmxi}{{\bm{x}_i}\xspace}
\newcommand{\bmxj}{{\bm{x}_j}\xspace}
\newcommand{\bmxpp}{{\bm{x}_p}\xspace}
\newcommand{\bmxqq}{{\bm{x}_q}\xspace}

\newcommand{\bmx}{\bm{x}\xspace}
\newcommand{\bmxp}{\bm{x'}\xspace}

\newcommand{\deltax}{\Delta\bm{x}\xspace}
\newcommand{\cPsf}[1]{{P_c\left(#1\right)\xspace}}
\newcommand{\aPsfSolo}{{P_a\xspace}}
\newcommand{\aPsf}[1]{{\aPsfSolo\left(#1\right)\xspace}}
\newcommand{\csig}[1]{{s\left(#1\right)\xspace}}
\newcommand{\ccsig}[1]{{s^*\left(#1\right)\xspace}}
\newcommand{\invivo}{\textit{in vivo}\xspace}

\newcommand{\invitro}{\textit{in vitro}\xspace}

\newcommand{\rbra}[1]{{\left({#1}\right)}}
\newcommand{\cbra}[1]{{\left\{{#1}\right\}}}
\newcommand{\abra}[1]{{\left[{#1}\right]}}
\newcommand{\mbra}[1]{{\left|{#1}\right|}}

\newcommand{\vdc}{s\rbra{\bmxp + \frac{\deltax}{2}}s^*\rbra{\bmxp - \frac{\deltax}{2}}}
\newcommand{\vdcinline}{s\rbra{\bmxp + \deltax/2}s^*\rbra{\bmxp - \deltax/2}}

\begin{document}

\title{Theoretical model for \enface optical coherence tomography imaging and its application to volumetric differential contrast imaging}

\author{Kiriko Tomita,\authormark{1} Shuichi Makita,\authormark{1} Naoki Fukutake,\authormark{2} Rion Morishita,\authormark{1} Ibrahim Abd El-Sadek,\authormark{1,3} Pradipta Mukherjee,\authormark{1} Antonia Lichtenegger,\authormark{1,4} Junya Tamaoki,\authormark{5} Lixuan Bian,\authormark{5} Makoto Kobayashi,\authormark{5} Tomoko Mori,\authormark{6} Satoshi Matsusaka,\authormark{6} and Yoshiaki Yasuno\authormark{1,*}}

\address{\authormark{1}Computational Optics Group, University of Tsukuba, Japan\\
	\authormark{2}Nikon Corporation, Japan\\
	\authormark{3}Department of Physics, Faculty of Science, Damietta University, Damietta, Egypt\\
	\authormark{4}Center for Medical Physics and Biomedical Engineering, Medical University of Vienna, Austria\\
	\authormark{5}Department of Molecular and Developmental Biology, Institute of Medicine, University of Tsukuba, Japan\\
	\authormark{6}Clinical Research and Regional Innovation, Faculty of Medicine, University of Tsukuba, Japan\\
}
\email{\authormark{*}yoshiaki.yasuno@cog-labs.org} 
\homepage{https://optics.bk.tsukuba.ac.jp/COG/} 

\begin{abstract}
	A new formulation of lateral imaging process of point-scanning optical coherence tomography (OCT) and a new differential contrast method designed by using this formulation are presented.
	The formulation is based on a mathematical sample model called the dispersed scatterer model (DSM), in which the sample is represented as a material with a spatially slowly varying refractive index and randomly distributed scatterers embedded in the material.
	It is shown that the formulation represents a meaningful OCT image and speckle as two independent mathematical quantities.
	The new differential contrast method is based on complex signal processing of OCT images, and the physical and numerical imaging processes of this method are jointly formulated using the same theoretical strategy as in the case of OCT.
	The formula shows that the method provides a spatially differential image of the sample structure.
	This differential imaging method is validated by measuring \invivo and \invitro samples.
\end{abstract}

\section{Introduction}
Several theoretical models of optical coherence tomography (OCT) \cite{Huang:91} have been studied to understand its image formation.
Because OCT is used in the biomedical field such as ophthalmology \cite{Curcio:11} and dermatology \cite{Sattler:13}, theoretical modeling is important for interpreting OCT images and clarifying what is visualized in the sample tissue.
Theoretical modeling is also useful for designing optimal OCT systems and developing image processing algorithms.

Several analytic and numerical models of OCT imaging have been developed.
After an early trial of OCT modeling \cite{Schmitt:93}, the extended Huygens-Fresnel formulation was introduced \cite{Schmitt:97, Thrane:00}.
In this model, propagation of the OCT probe beam in a tissue, i.e., an inhomogeneous medium, is modeled by extended Huygens-Fresnel formulation. Numerical simulations based on this model can recapitulate the forward multiple scatting in the tissue.
Monte-Carlo methods were also widely used to numerically emulate the OCT imaging \cite{Smithies_1998,Yao_1999}.
Although Monte-Carlo methods are nondeterministic and computationally expensive in general, they can be applied to wide varieties of imaging models, conditions, and samples.
The interaction between the probe beam and the sample can be also accurately computed by using Maxwell's equation \cite{Munro:15, Brenner:16}.
An extension of this method, the pseudo-spectral time domain method, enables the realistic simulation of three-dimensional OCT imaging \cite{Munro:16}.
Recently, the speckle was mathematically described as the interference of light backscattered by multiple scatterers in a resolution volume \cite{KCZhou2021AOP}.

Aforementioned approaches aimed to recapture accurate and realistic OCT signals.
By contrast, we aim to establish a simplified theoretical framework that enables the intuitive understanding of the point-scanning OCT imaging process and the interpretation of an \enface OCT image.
In this framework, we model a sample as a spatially slowly varying refractive index in which scatterers are randomly dispersed (Section \ref{sec:DSM}).
The density of the scatterers is also spatially slowly varying.
This representation of the sample is denoted as a ``dispersed scatterer model'' (DSM).
We formulate the OCT imaging process using the DSM (Section \ref{sec:OctFormulation}).
This formulation identifies the relationship between the sample structure, that is, the slowly varying refractive index distribution and the scatterer density distribution, and the \enface OCT signal.
In this formulation, a meaningful OCT signal and speckle appear as different terms; hence, it provides a comprehensive understanding of the \enface OCT image formation process.

By actively using this theoretical framework, we design and experimentally demonstrate an OCT-based volumetric differential contrast (VDC) imaging method (Section \ref{sec:VDC}).
We implement this method as the numerical signal processing flow of complex \enface OCT images and provide differential contrast images similar to those of differential interference contrast microscopy (DIC) \cite{Allen:69}.
Similar to DIC, VDC highlights the micro-structure of a sample, but unlike DIC, VDC can be volumetric.

The DSM-based framework can also explain phenomena that occur in OCT-based imaging methods but have not been theoretically well explained.
An example is artifacts found in A-scan-wise Doppler OCT images caused by the structure of superior tissue.
Another example relates to the specific imaging condition required for point-scanning computational multi-directional OCT \cite{Oida_2021}.
In experiments, it was found that defocus is mandatory to achieve this method \cite{Oida_bios}, but this necessity has not been explained in theory.
We analyze these issues using the DSM-based framework in Section \ref{sec:discussion}.

\section{Dispersed scatterer model (DSM)}
\label{sec:DSM}
In our theoretical framework, we model a sample to be measured as a spatially slowly varying refractive index distribution in which scatterers are spatially randomly dispersed.
We consider the density of the scatterers to be also spatially slowly varying.
We refer to this theoretical model of the sample as a DSM.

\begin{figure}
	\centering\includegraphics[width=9cm]{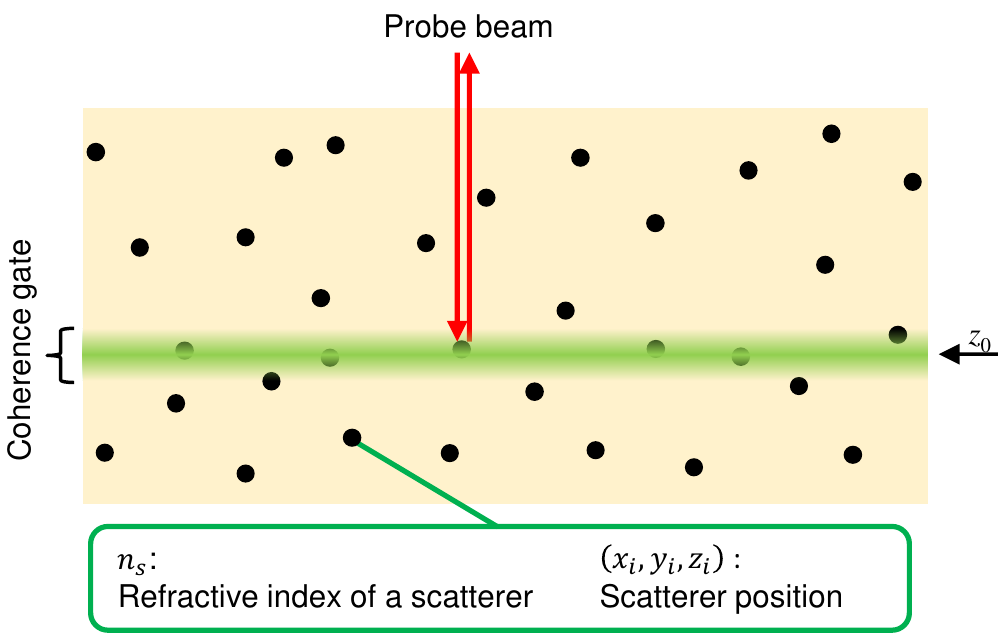}
	\caption{
		The schematic model of the sample, i.e., the dispersed scatterer model (DSM).
		The sample is represented as a spatially slowly varying refractive index $n(x,y,z)$ in which infinitely small scatterers (black dots) are randomly dispersed.
		All the scatterers are assumed to have the same refractive index of $n_s$.
		The green region represents the coherence gate of OCT and only the scatterers within the coherence gate contribute to OCT imaging.
		The probe beam is incident on the sample from the top and only lights backscattered toward the incident direction contribute to imaging.
	}
	\label{fig:DSM}
\end{figure}

Figure \ref{fig:DSM} schematically illustrates the DSM, where $(x,y,z)$ is the position in the sample as $x$ and $y$ are the lateral positions, and $z$ is the depth position. 
The spatially slowly varying refractive index is represented by $n(x,y,z)$, in which small scatterers are dispersed (i.e., embedded), where the scatterers are assumed to be significantly smaller than the spatial variation of $n(x,y,z)$.
The position of the $i$-th scatterer is $(x_i, y_i, z_i)$ and all the scatterers have an identical refractive index of $n_s$.
The probe beam is incident from above of the sample, and the green region in the figure represents the coherence gate.
In our formulations for OCT (Section \ref{sec:OctFormulation}) and VDC (Section \ref{sec:VDC}), we consider \enface images in which scattering only within this coherence gate contributes to imaging.

\section{Formulation of OCT imaging}
In this section, we formulate the image formation process of standard scanning OCT imaging using DSM.
\label{sec:OctFormulation}

\subsection{Complex OCT signal} 
We assume that a Gaussian beam is incident on the sample.
Considering that $N$ scatterers are involved in imaging, i.e., $N$ scatterers exist in the coherence gate, a complex \enface OCT signal becomes
\begin{equation}
	s(\bmxp; z'_0)
	= \sum_{i=1}^{N} B(\bmxi, z_0) e^{i\phi_i} \cPsf{\bmxp - \bmxi; z'_0},
	\label{eq:ComplexOCTSignal}
\end{equation}
where $\bm{x}=(x,y)$ and $\bmxp=(x',y')$ are the two-dimensional lateral positions on a sample and image, respectively.
Here the contribution from each scatterer is that of single scattering. Namely, the first-order Born approximation has been employed.
Throughout this paper, we use $\bm{x}$ and $z$ to represent the position in the object (sample) space, and $\bmxp$ and $z'$ to represent the position in the image space.
$z_0'$ is the depth position in the image corresponding to the center depth of coherence gate $z_0$ in the sample.
The magnification between the sample and the image is assumed to be 1.
$\bmxi=(x_i,y_i)$ is the lateral position of the $i-$th scatterer.
$B(\bm{x}, z_0)$ is a quantity defined by the slowly varying refractive index distribution of the medium, $n(x,y,z)$, and the refractive index of the embedded scatterers, $n_s$.
And hence, $B$ is also spatially slowly varying as is $n$.
It is noteworthy that the overall profile of $B(\bm{x}, z_0)$ is defined solely by the profile of the spatially slowly varying refractive index distribution $n(x,y,z)$.
The spatial inhomogeneity of the dispersed scatterers is accounted for later when we derive the continuous form of the imaging formula in Section \ref{sec:continuousIntensity}.
For the case of sufficiently small scatterers, $B(\bm{x}, z_0)$ can be considered as the scattering potential 
$B(\bm{x}, z_0) \simeq \cbra{2\pi n(\bm{x}, z_0)/\lambda_0} \left[\left\{n_s/n(\bm{x}, z_0)\right\}^2-1\right]$, where $\lambda_0$ is the central wavelength of the probe beam \cite{BornWolf2019Chap13}.

$\cPsf{\bmx; z'_0}$ is a complex point spread function (PSF) at depth $z'_0$.
$\phi_i$ is the phase offset caused by the depth position of the $i$-th scatterer with respect to the central depth of the coherence gate $z_0$:
\begin{equation}
	\phi_i = k_0\, n\, 2(z_i-z_0),
\end{equation}
where $k_0 = 2 \pi/\lambda_0$ is the wavenumber as $\lambda_0$ is the center wavelength of the probe beam and $n$ is the refractive index of the sample around the scatterers.
Note that we assume that $B$ is homogeneous along the depth within the coherence gate; hence, the variation of the scatterer's depth position does not affect $B$ but only appears as phase offset. 

The complex PSF $\cPsf{\bmx; z'_0}$ is further broken down as
\begin{equation}
	\label{eq:cPsf}
	\cPsf{\bmx; z'_0} \equiv \aPsf{\bmx; z'_0}\,  
	e^{i \varphi \left(\bmx ; z_0 \right) },
\end{equation}
where $P_a \equiv \mbra{P_c}$ is the amplitude of the PSF and is typically Gaussian in the lateral direction, and $\varphi$ is the phase of the complex PSF.
We discuss a more practical shape for the complex PSF in Section \ref{sec:GaussianPSF}.

By substituting Eq.\@ (\ref{eq:cPsf}) into Eq.\@ (\ref{eq:ComplexOCTSignal}), the complex OCT signal $\csig{\bmxp; z'_0}$ becomes
\begin{equation}
	\csig{\bmxp; z'_0}
	= \sum_{i=1}^{N} B(\bmxi, z_0) e^{i\phi_i} \aPsf{\bmxp - \bmxi; z'_0}\, 
	e^{i \varphi \left( \bmxp - \bmxi ; z_0 \right) }.
	\label{eq:ComplexOCTSignalBD}
\end{equation}
As shown in Eq.\@ (\ref{eq:ComplexOCTSignalBD}), the contribution from each scatterer is weighted by the amplitude of the complex PSF, $P_a$, and its phase is modulated by both the depth position of the scatterer and spatial phase pattern of the PSF.

\subsection{Intensity-OCT signal in discrete form}
The intensity of OCT is the absolute square of the complex OCT signal, that is, the product of the complex OCT signal and its conjugate.
From Eq.\@ (\ref{eq:ComplexOCTSignalBD}), the \enface OCT intensity $I\left(\bmxp;z_0\right)$ becomes
\begin{align}
	I\left( \bmxp; z'_0 \right)
	&= \csig{\bmxp; z'_0} \ccsig{\bmxp; z'_0} \nonumber\\
	&= \left[ \sum_{i=1}^{N} B(\bmxi, z_0) e^{i\phi_i} 
	\aPsf{\bmxp - \bmxi; z'_0}\, e^{i \varphi \left( \bmxp - \bmxi \right) }\right]
	\nonumber \\
	& \quad \times \left[ \sum_{j=1}^{N} B(\bmxj, z_0) e^{-i\phi_j} \aPsf{\bmxp - \bmxj; z'_0}\, e^{-i \varphi \left( \bmxp - \bmxj \right) }\right] 
	\nonumber \\
	& = \sum_{i=1}^{N}\sum_{j=1}^{N}  
	B(\bmxi) B(\bmxj) e^{i(\phi_i-\phi_j)} \aPsf{\bmxp - \bmxi} 
	\aPsf{\bmxp - \bmxj} 
	e^{i \left[\varphi \left( \bmxp - \bmxi \right) - \varphi \left( \bmxp - \bmxj \right) \right]},
	\label{eq:OCTInt_sum}
\end{align}	
where $\bmxj=(x_j,y_j)$ is the position of the $j$-th scatterer in the sample.
On the last line of the equation, we did not explicitly write $z_0$ and $z'_0$ for simplicity.
We also omitted them in the later part of the paper if it did not cause confusion.

\begin{figure}
	\centering\includegraphics[width=13cm]{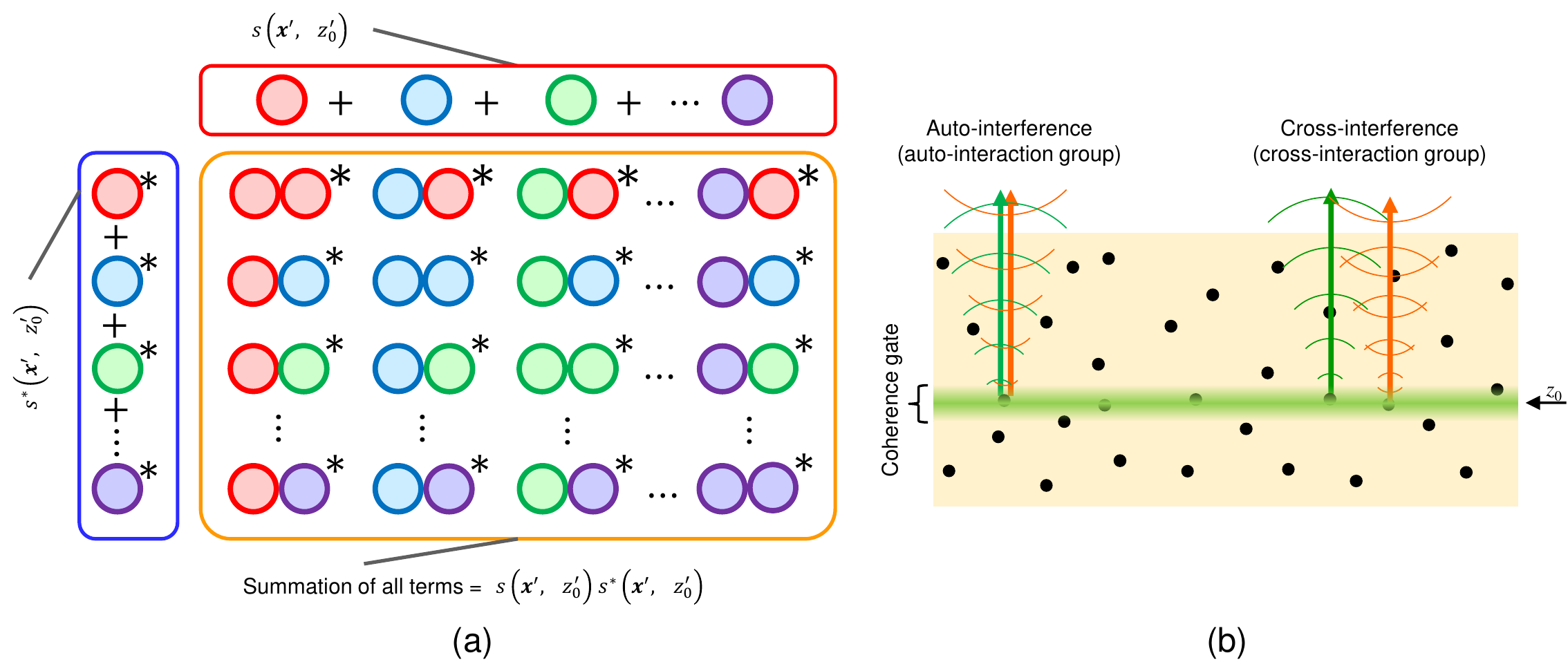}
	\caption{
        (a) Schematic of the formulation of OCT intensity. Each circle represents each term of the equation.
        The same color indicates the same term and the asterisk represents the complex conjugate. The red box consists of the terms on the second line of Eq.\@ (\ref{eq:OCTInt_sum}) and the blue box corresponds to the terms on the third line of Eq.\@ (\ref{eq:OCTInt_sum}). The orange box represents the terms on the fourth line of Eq.\@ (\ref{eq:OCTInt_sum}).
        (b) Physical depiction of the numerical interactions between two signals contributed by each pair of scatterers, which are mathematically described by Eq.\@ (\ref{eq:OctIntensityGrouped}).
        The interactions are classified into two groups: auto-interaction (AI) and cross-interaction (CI) groups.
        AI corresponds to the numerical interference of the electric field scattered from a scatterer and the conjugate field from the same scatterer; hence, it corresponds to the backscattered intensity from the scatterer.
        CI corresponds to the numerical interference of the electric field from a scatterer and the conjugate field from another scatterer; hence, is a phasor.
        AI and CI are also referred to as auto-interference and cross-interference, respectively.
	}
	\label{fig:AIandCI}
\end{figure}
Eq.\@ (\ref{eq:OCTInt_sum}) is schematically depicted in Fig.\@ \ref{fig:AIandCI}(a).
Each circle represents each term of the equation, that is, the contribution from each scatterer in the sample.
The same color indicates the same term, that is, the contribution from the same scatterer, and the asterisk represents the complex conjugate.
The red box consists of the terms in $\csig{\bmxp, z'_0}$, that is, the second line of Eq.\@ (\ref{eq:OCTInt_sum}) and the blue box represents $\ccsig{\bmxp, z'_0}$, which corresponds to the third line of Eq.\@ (\ref{eq:OCTInt_sum}).
The orange box represents interactions among the terms in $\csig{\bmxp, z'_0}$ and $\ccsig{\bmxp, z'_0}$, that is, the terms on the fourth line of Eq.\@ (\ref{eq:OCTInt_sum}).
We use this depiction in the interpretation of the equation in the next paragraph.

The terms of Eq.\@ (\ref{eq:OCTInt_sum}) can be classified into two groups for $i = j$ and $i \neq j$ as
\begin{multline}
	\label{eq:OctIntensityGrouped}
	I\left( \bmxp\right) 
	= 
	\color{darkblue}
	\sum_{i=1}^{N} B^2(\bmxi) P_a^2 \left(\bmxp - \bmxi \right) \\  
	+ 
	\color{auburn}
	\underset{(i \neq j)}{\sum_{i=1}^{N}\sum_{j=1}^{N}}  
	B(\bmxi) B(\bmxj) \color{darkred} e^{i(\phi_i-\phi_j)} \aPsf{\bmxp - \bmxi} 
	\aPsf{\bmxp - \bmxj} 
	e^{i \left[\varphi \left( \bmxp - \bmxi \right) - \varphi \left( \bmxp - \bmxj \right) \right]}
	\color{black}
	.
\end{multline}
This equation is the representation of the OCT intensity signal in the discrete form of our theoretical framework.
In this formulation, each term in the first group (the first line of the equation, for $i = j$, written in blue) is the numerical interaction between contributions from the same scatterer; hence, we refer to this group as an auto-interaction (AI) group.
By contrast, each term in the second group (the second line, for $i \neq j$, written in brown and red) is the numerical interaction between contributions from different scatterers.
We refer to this group as a cross-interaction (CI) group.
We classified CI into brown and red parts, and this classification is used in later sections.
The terms of the AI group correspond to the diagonal terms in the orange box in Fig.\@ \ref{fig:AIandCI}(a), whereas those of the CI group are the off-diagonal terms.

It is noteworthy that these interactions can be considered as the numerical interferences of scattered electric fields from the scatterers as depicted in Fig.\@ \ref{fig:AIandCI}(b).
The AI term corresponds to interference between an electric field from a scatterer and the complex conjugate of the same electric field. 
Hence, each term in the AI group represents the squared intensity of scattered light from each scatterer and the AI group is the incoherence summation of these scattered light intensities.
By contrast, each term in the CI group is the numerical interference between the scattered complex field from a scatterer and the conjugate complex field from another scatterer, where two scatterers are sufficiently close to be included in the complex PSF.
Note that each term in the CI group is a phasor with a practically random phase and each term in the CI group has its complex conjugate pair, as can be schematically seen from the diagonal symmetricity of the orange box in Fig.\@ \ref{fig:AIandCI}(a).
Because of the random phase of each term, the summation of a term and its complex conjugate becomes a random real value.
This suggests that the CI group causes a random signal, i.e., speckle, as we discuss in detail in Section \ref{sec:interpretationOfIntensity}.

\subsection{Intensity-OCT signal in continuous form}
\label{sec:continuousIntensity}
In Eq.\@ (\ref{eq:OctIntensityGrouped}), the OCT intensity is expressed using the contributions from discrete scatterers; hence, the equation is also in discrete form. 
In this subsection, we derive a continuous-form representation of the OCT intensity, which is a convolution-based conventional image formation equation, and hence convenient to interpret.

Using the Dirac delta function, the intensity-OCT signal in discrete form [Eq.\@ (\ref{eq:OctIntensityGrouped})] can be rewritten as 
\begin{multline}
	\label{eq:OctInterContinuous}
	I(\bmxp)
	= 
	\color{darkblue}
	\iint \sum_{i=1}^{N} \delta \rbra{\bmxpp - \bmxi}
        B^2(\bmxpp) P_a^2(\bmxp -\bmxpp)  d\bmxpp \\  
         +
         \color{auburn}
         \biggl[ \iiiint  \sum_{i=1}^{N} \sum_{j=1}^{N} 
         \delta \rbra{\bmxpp - \bmxi, \bmxqq - \bmxj}
         B(\bmxpp) B(\bmxqq) \color{darkred} e^{i(\phi_{pq})} \\
	\color{darkred}
	\times \aPsf{\bmxp - \bmxpp} \aPsf{\bmxp - \bmxqq} 
	e^{i \left[\varphi \left( \bmxp - \bmxpp \right) - \varphi \left( \bmxp -\bmxqq \right) \right]} 
	\color{auburn} d\bmxpp d\bmxqq \\
	\color{auburn}
	 -\iint \sum_{i=1}^{N} \delta \rbra{\bmxpp - \bmxi}
        B^2(\bmxpp) P_a^2(\bmxp - \bmxpp)  d\bmxpp \biggr] \color{black},
\end{multline}
where $\bmxpp$ and $\bmxqq$ are variables in real physical space in the sample.
Each of the double integral is for $x$ and $y$ components of the vectorized integration variable $\bmxpp$, while the quadruple integral is for two vectorized integration variables $\bmxpp$ and $\bmxqq$.
The intervals of all the integrals are from $-\infty$ to $+\infty$.
Hereafter, the integral intervals of all equations are from $-\infty$ to $+\infty$; we omit them for simplicity.
$\phi_{pq}$ is a phase that is originated from $\phi_i-\phi_j$ in Eq.\@ (\ref{eq:OctIntensityGrouped}).
Namely, $\phi_{pq}$ is the phase difference between the contributions of $i$-th and $j$-th scatterer and equals to $\phi_i-\phi_j$.
Similar to $\phi_i-\phi_j$, $\phi_{pq}$ is unpredictable and random in practice, and $\phi_{pq} = -\phi_{qp}$.
Note that, in the second line of Eq.\@ (\ref{eq:OctInterContinuous}), we can consider that $\bmxpp$ and $\bmxqq$ respectively denote the positions of $i$-th and $j$-th scatterers due to the Dirac delta function.
The first line of Eq.\@ (\ref{eq:OctInterContinuous}) (blue) corresponds to the AI group [the first line of Eq.\@ (\ref{eq:OctIntensityGrouped})] and the second to fourth lines of Eq.\@ (\ref{eq:OctInterContinuous}) (brown and red) correspond to the CI group [the second line of Eq.\@ (\ref{eq:OctIntensityGrouped})].
Note that the summation in CI of Eq.\@ (\ref{eq:OctIntensityGrouped}) is for $i \neq j$, whereas we removed this condition in Eq.\@ (\ref{eq:OctInterContinuous}).
Alternatively, we subtract the fourth line, which is identical to the first line corresponding to the case of $i=j$.
This removal of condition was necessary to convert the discrete equation [Eq.\@ (\ref{eq:OctIntensityGrouped})] into the continuous equation [Eq.\@ (\ref{eq:OctInterContinuous})] because the integral cannot properly account for such a condition.

We introduce functions $D$ and $D'$ as
\begin{equation}
    D \rbra{\bmxpp} \equiv \sum_{i=1}^{N} \delta \rbra{\bmxpp - \bmxi},
    \label{eq:D}
\end{equation}
\begin{equation}
    D' \rbra{\bmxpp, \bmxqq} \equiv \sum_{i=1}^{N} \sum_{j=1}^{N} 
    \delta \rbra{\bmxpp - \bmxi, \bmxqq - \bmxj},
    \label{eq:D'}
\end{equation}
which are two- and four-dimensional binary maps of the scatterers' positions, respectively.
When the scatterer density is high enough to be considered as continuously distributed, that is, $N$ is sufficiently large, $D$ and $D'$ can be considered as non-binary scatterer density maps in two- and four-dimensional spaces, respectively.
Specifically, $D$ is the scatterer density map of the sample.

By substituting Eqs.\@ (\ref{eq:D}) and (\ref{eq:D'}) into Eq.\@ (\ref{eq:OctInterContinuous}), we obtain the continuous form of the equation:
\begin{multline}
	\label{eq:OCTInt_integral}
	I\rbra{\bmxp} 
	=
	\color{darkblue}
	\iint D(\bmxpp) B^2(\bmxpp) P_a^2 \rbra{\bmxp - \bmxpp} d \bmxpp \\
	+
	\color{auburn}
	\left[ \iiiint D'(\bmxpp, \bmxqq) B(\bmxpp) B(\bmxqq) \right.\\
	\color{darkred}
	\left. e^{i(\phi_{pq})} \right. 
	\color{darkred}
	\aPsf{\bmxp - \bmxpp} \aPsf{\bmxp - \bmxqq} 
	e^{i \left[\varphi \left( \bmxp - \bmxpp \right) 
		- \varphi \left(\bmxp - \bmxqq \right)     \right]} \color{auburn} d \bmxpp d \bmxqq \\
	\color{auburn}
	\left. - \iint D(\bmxpp) B^2(\bmxpp) 
	P_a^2 \rbra{\bmxp - \bmxpp} d \bmxpp \right] \color{black}.
\end{multline}
The first integral (i.e., the AI part) of Eq.\@ (\ref{eq:OCTInt_integral}) (blue) is ``an imaging equation of the intensity-OCT image'' in continuous form.
According to this equation, the object to be imaged in the intensity-OCT image is the product of the squared scattering potential $B^2$ and the scatterer density $D$, and the PSF of intensity-OCT imaging is $P_a^2$.
A more detailed interpretation of this intensity-OCT imaging formula is provided in the following subsections (Sections \ref{sec:GaussianPSF} and \ref{sec:interpretationOfIntensity}). 

\subsection{Intensity-OCT imaging with Gaussian PSF}
\label{sec:GaussianPSF}
The property of intensity-OCT imaging becomes clearer if we assume a complex Gaussian PSF with defocus in the continuous OCT imaging formula [Eq.\@ (\ref{eq:OCTInt_integral})].
The complex Gaussian PSF is obtained from a Gaussian probe beam, and the Gaussian probe beam is typical for a single-mode-fiber-based point-scanning OCT.
According to the formulation by Ralston \etal \cite{Ralston:05}, the Gaussian probe beam in the paraxial approximation becomes \begin{equation}
	\label{eq:GaussianBeam}
	G(\bmx, z;z_0) = \frac{w_0}{w(z;z_0)}
	\exp\left[{- \frac{\bmx^2}{w^2(z;z_0)}}\right]
	\exp \left[-i \left\{n k_0 (z-z_0) + \frac{n k_0 \bmx^2}{2R(z;z_0)} - \psi(z;z_0) \right\}\right],
\end{equation}
where $z_0$ is the depth position of the focus, $w(z;z_0)$ is the radius of the probe beam at depth $z$, $w_0 = w(z = z_0; z_0)$ is the beam radius at the in-focus depth, that is, the beam waist radius, and $n$ is the refractive index of the sample around the position of interest.
The second term in the second exponential represents the quadratic phase induced by defocus and $R(z;z_0)$ is the phase curvature.
$\bmx^2$ represents $\bmx \cdot \bmx$.
Here after the squared vector represents the inner product of a vector with itself.
In the third term of the second exponential, $\psi(z;z_0)$ represents the Gouy phase.

The beam waist radius $w_0$, the radius of the probe beam $w$, and the phase curvature $R$ are
\begin{equation}
	w_0 = \frac{4 f}{n \, \Phi \, k_0},
	\label{eq:w0}
\end{equation}
\begin{equation}
	w(z;z_0) = w_0 \sqrt{1 + \cbra{\frac{2 (z-z_0)}{n^2 \, k_0 \, w_0^2}}^2},
	\label{eq:w}
\end{equation}
\begin{equation}
	R(z;z_0)= \rbra{z-z_0} \abra{1 + \cbra{\frac{n^2 \, k_0 \, w_0^2}{2 (z-z_0)}}^2 },
	\label{eq:R}
\end{equation}
where $f$ is the focal length of the objective and $\Phi$ is the $1/e^2$-diameter of the probe beam incident on the objective.

Because OCT is a reflective imaging modality, and the illumination and collection paths share the same optics, the complex PSF $\cPsf{\bmx, z; z_0}$ is obtained by multiplying $G(\bmx, z;z_0)$ by itself as
\begin{multline}
	\cPsf{\bmx,z ;z_0} = G(\bmx, z;z_0)\, G(\bmx, z;z_0) \\
	= \frac{w_0^2}{w^2(z;z_0)}
	\exp \left[ - \frac{2 \bmx^2}{w^2(z;z_0)} \right]
	\exp \left[-i \left\{2 n k_0 (z-z_0) + \frac{n k_0 \bmx^2}{R(z;z_0)} - 2 \psi(z;z_0) \right\}\right].
	\label{eq:PSF}
\end{multline}
On the first line, each $G(\bmx, z;z_0)$ is for probe beam illumination and collection, respectively.

By comparing this complex PSF and that of Eq.\@ (\ref{eq:cPsf}), we find that 
\begin{equation}
	\label{eq:GaussPa}
	\aPsf{\bmx, z';z'_0}=
	\frac{w_0^2}{w^2(z';z'_0)}
	\exp \left[ - \frac{2 \bmx^2}{w^2(z';z'_0)} \right]
\end{equation}
and
\begin{equation}
	\label{eq:GaussPhi}
	\varphi(\bmx, z; z_0) = - 2 n k_0 (z-z_0) - \frac{n k_0}{R(z;z_0)} \bmx^2 + 2 \psi (z;z_0).
\end{equation}

By substituting this Gaussian complex PSF, or equally $P_a$ and $\varphi$, into Eq.\@ (\ref{eq:OctIntensityGrouped}), we obtain the discrete-form intensity-OCT imaging equation:
\begin{multline}
	I(\bmxp) =
	\color{darkblue}
	\frac{w_0^4}{w^4} \sum_{i=1}^{N} B^2(\bmxi) 
	\exp\abra{- \frac{4}{w^2} \left(\bmxp - \bmxi \right)^2} \\
	+ 
	\color{auburn}
	\frac{w_0^4}{w^4} \underset{(i \neq j)}{\sum_{i=1}^{N}\sum_{j=1}^{N}}
	B(\bmxi) B(\bmxj)
	\color{darkred}
	\exp\abra{- \frac{(\bmxi-\bmxj)^2}{w^2}}
        \exp i \abra{\phi_i - \phi_j}
         \\
         \color{darkred}
        \times \exp\abra{- \frac{4}{w^2} \rbra{\bmxp - \frac{\bmxi+\bmxj}{2}}^2}
        \exp i \abra{\frac{2 n k_0}{R} \rbra{\bmxi - \bmxj }
        \cdot \rbra{\bmxp - \frac{\bmxi + \bmxj}{2}}} \color{black}.
	\label{OctInt_Discrete_PSF}
\end{multline}
We omit depth positions for simplicity.
The first line of this equation (blue) corresponds the AI group, while second and third lines (brown and red) correspond to the CI group.
Using the symmetricity of $\bmxi$ and $\bmxj$, the second summation part can be rewritten in a real form from Eq.\@ (\ref{OctInt_Discrete_PSF}) as
\begin{multline}
	I(\bmxp) =
	\color{darkblue}
	\frac{w_0^4}{w^4} \sum_{i=1}^{N} B^2(\bmxi) 
	\exp\abra{- \frac{4}{w^2} \left(\bmxp - \bmxi \right)^2} \\
	+ 
	\color{auburn}
	\frac{2 w_0^4}{w^4} \underset{(i > j)}{\sum_{i=1}^{N}\sum_{j=1}^{N}}
	B(\bmxi) B(\bmxj)
	\color{darkred}
	\exp\abra{- \frac{(\bmxi-\bmxj)^2}{w^2}}
	\exp\abra{- \frac{4}{w^2} \rbra{\bmxp - \frac{\bmxi+\bmxj}{2}}^2} \\
	\color{darkred}
	\times \cos\abra{(\phi_i - \phi_j) + \frac{2 n k_0}{R} 
        \rbra{\bmxi - \bmxj} \cdot \rbra{\bmxp - \frac{\bmxi + \bmxj}{2}}
        } \color{black}.
	\label{eq:IntensityPsfDiscReal}
	\\
\end{multline}

By applying the same manipulations used to convert Eq.\@ (\ref{eq:OctIntensityGrouped}) to Eq.\@ (\ref{eq:OCTInt_integral}), which we described in Section \ref{sec:continuousIntensity}, we obtain the continuous form of the intensity-OCT equation with the Gaussian PSF with defocus from Eq.\@ (\ref{OctInt_Discrete_PSF}) as
\begin{multline}
	I(\bmxp)
	= 
	\color{darkblue}
	\frac{w_0^4}{w^4} 
	\iint D(\bmxpp) B^2(\bmxqq) 
	\exp\abra{- \frac{4}{w^2} \left(\bmxp - \bmxpp \right)^2} d \bmxpp \\
	+ 
	\color{auburn}
	\frac{w_0^4}{w^4}
	\iiiint D'(\bmxpp,\bmxqq)  B(\bmxpp) B(\bmxqq)
		\color{darkred}
        \exp\abra{- \frac{(\bmxpp-\bmxqq)^2}{w^2}} 
        e^{i \phi_{pq}} \\
    \color{darkred}
	\times 
	\exp\abra{- \frac{4}{w^2} \rbra{\bmxp - \frac{\bmxpp+\bmxqq}{2}}^2} 
	\exp i \abra{\frac{2 n k_0}{R} 
        \rbra{\bmxpp - \bmxqq} \cdot \rbra{\bmxp - \frac{\bmxpp + \bmxqq}{2}}}
        \color{auburn} d\bmxpp d\bmxqq\\
    \color{auburn} 
	-\mathrm{the\,\,\, first\,\,\, line \,\,\, of \,\,\, R.H.S.},
	\label{eq:Intensity_PSF}
\end{multline}
where R.H.S. means ``the right hand side.''
The last line, which is identical to the AI group (blue), was subtracted from the second integral, so that the last three lines of the equation (brown and red) represent the CI group.
This equation is the continuous-form imaging equation of an intensity-OCT image with a complex Gaussian PSF with defocus.
The first line and other lines correspond to AI and CI, respectively.

Using the symmetricity of $\bmxpp$ and $\bmxqq$, and $\phi_{pq} = - \phi_{qp}$, the second integral of Eq.\@ (\ref{eq:Intensity_PSF}) is rewritten in a real form as
\begin{multline}
	I(\bmxp)
	= \color{darkblue} 
	\frac{w_0^4}{w^4} 
	\iint D(\bmxpp) B^2(\bmxpp) 
	\exp\abra{- \frac{4}{w^2} \left(\bmxp - \bmxpp \right)^2} d \bmxpp \\
	+ \color{auburn} 
	\frac{w_0^4}{w^4} 
	\iiiint D'(\bmxpp,\bmxqq)  B(\bmxpp) B(\bmxqq)
	\color{darkred}
	\exp\abra{- \frac{(\bmxpp-\bmxqq)^2}{w^2}}
	\exp\abra{- \frac{4}{w^2} \rbra{\bmxp - \frac{\bmxpp+\bmxqq}{2}}^2} \\
	\color{darkred}
	\times
	\cos\abra{\phi_{pq} + \frac{2 n k_0}{R} 
		\rbra{\bmxpp - \bmxqq} \cdot \rbra{\bmxp - \frac{\bmxpp + \bmxqq}{2}}}  
	\color{auburn} d \bmxpp d \bmxqq\\
	\color{auburn}
	-\mathrm{the\,\,\, first\,\,\, line \,\,\, of \,\,\, R.H.S.}
	\label{eq:IntensityPsfContReal}
\end{multline}

\subsection{Interpretation of OCT intensity}
\label{sec:interpretationOfIntensity}

Eqs.\@ (\ref{eq:OctIntensityGrouped}), (\ref{eq:OCTInt_integral}), (\ref{eq:IntensityPsfDiscReal}), and (\ref{eq:IntensityPsfContReal}) provide an intuitive understanding of the intensity-OCT image.
All the equations suggest that the OCT intensity consists of AI [i.e., the first summation or integral of each equation (blue) of Eqs.\@ (\ref{eq:OctIntensityGrouped}), (\ref{eq:OCTInt_integral}), (\ref{eq:IntensityPsfDiscReal}), and (\ref{eq:IntensityPsfContReal})], and CI (i.e., the residual terms, brown and red). 
AI forms a meaningful image, as we discuss in this paragraph, whereas CI corresponds to speckle, which we discuss in the next paragraph.
As can be seen in the equations, for the intensity-OCT image formed by AI, the object to be measured is $B^2(\bmxi)$ for discrete forms and $D(\bmx)B^2(\bmx)$ for continuous forms.
Additionally, it is noteworthy that this intensity-OCT imaging explicitly has a real- and positive-valued PSF: $\aPsfSolo^2$.
We found that the AIs of the continuous-form equations [Eqs.\@ (\ref{eq:OCTInt_integral}) and (\ref{eq:IntensityPsfContReal})] are the convolution of the real-positive PSF and the object to be measured.
And hence, the AI is equivalent to the incoherent image of the object.

In contrast to AI, CI does not have an explicit PSF.
To intuitively understand the image property of CI, we split the product parts inside the summation or integral into two parts: the object to be measured ($B^2(\bmxi)$ or $D(\bmx)B^2(\bmx)$) and the residual part (written in red).
We refer to the residual part as ``pseudo-PSF'' because it can be considered as the PSF for each scatterer pair.
The pseudo-PSF varies for each combination of scatterers; hence, CI does not form a meaningful image.
Namely, CI represents speckle.
Specifically for the intensity-OCT imaging under a Gaussian complex PSF [Eqs.\@ (\ref{eq:IntensityPsfDiscReal}) and (\ref{eq:IntensityPsfContReal})], the pseudo-PSF is found to have the following characteristics.
First, the pseudo-PSF is the product of the Gaussian envelope (the second exponential parts  in CI) and a cosine wave.
Second, the Gaussian envelope has the same width as the explicit PSF of AI (the exponential part of AI).
It is consistent with a well-known fact; speckle size is a function of the OCT resolution \cite{Schmitt:97, Gossage_2006}.
Third, the center of the Gaussian is located at the center of the two scatterers of each scatterer pair;
that is, the center of the Gaussian envelope walks around for each scatterer pair and is practically unpredictable and random.
Fourth, the frequency and phase offset, that is, the shift of the cosine wave, depends on the positions of the scatterers of the scatterer pair; hence, they are practically unpredictable and random.

These formulations indicate that the intensity-OCT image is a summation of a speckle-free meaningful image, i.e., the incoherent image of the object, which corresponds to AI, and a random speckle image corresponding to CI.


\section{Volumetric differential contrast (VDC) imaging}
\label{sec:VDC}

\subsection{Theory}
In this section, we present a new OCT-based imaging method called volumetric differential contrast (VDC) imaging, which we design using the DSM and a similar theoretical framework presented in Section \ref{sec:OctFormulation}.
This is an image processing method that manipulates the \enface OCT signal and can provide a spatially differentiated image of a sample that highlights the sample structure as contrast, similar to DIC.
In contrast to standard DIC, VDC is volumetric, that is, it provides a spatially differentiated image at each depth of a sample.

An \enface VDC image is obtained by multiplying an \enface complex OCT by its digitally shifted complex conjugate and then extracting the imaginary part.
We describe the formulation of VDC in this section.

In the first step of an \enface VDC computation, we compute the product of an \enface complex OCT signal and the digitally laterally shifted complex conjugate of this signal.
Using the same theoretical flow that we used to obtain Eq.\@ (\ref{eq:OCTInt_integral}), this product, which is an \enface complex signal, is written as
\begin{multline}
	\vdc 
	= 
	\color{darkblue}
	\iint D(\bmxpp) B^2(\bmxpp) 
	\aPsf{\bmxp - \bmxpp+\frac{\deltax}{2}}
	\aPsf{\bmxp - \bmxpp - \frac{\deltax}{2}}\\
	\color{darkblue}
	\times \exp {i \left[\varphi \left(\bmxp - \bmxpp+\frac{\deltax}{2} \right)    - \varphi \left(\bmxp - \bmxpp-\frac{\deltax}{2} \right) \right]} d \bmxpp
	\\  
	+
	\color{auburn}
	\iiiint D'(\bmxpp,\bmxqq)  B(\bmxpp)
	B(\bmxqq) 
	\color{darkred}
	e^{i (\phi_{pq})} \aPsf{\bmxp - \bmxpp+\frac{\deltax}{2}}
	\aPsf{\bmxp - \bmxqq - \frac{\deltax}{2}} \\
	\qquad \color{darkred} \times 
	\exp{i \left[ \varphi \left(\bmxp - \bmxpp+\frac{\deltax}{2} \right)
		-  \varphi \left(\bmxp - \bmxqq-\frac{\deltax}{2} \right) \right]} \color{auburn} d\bmxpp d\bmxqq \\
	\color{auburn} -\mathrm{the\,\,\, first\,\,\, integral \,\,\, of \,\,\, R.H.S.}\color{black},
	\label{eq:s1s2_cont}
\end{multline}
where $\deltax$ is the digital shift amount between the two signals, typically one or two lateral pixels, and should be smaller than the spot size of the probe beam.
The letter colors corresponds to those in the previous equations.
The detailed derivation of this equation is in the Appendix.

By assuming the Gaussian complex PSF described in Section \ref{sec:GaussianPSF}, that is, by substituting Eqs.\@ (\ref{eq:GaussPa}) and (\ref{eq:GaussPhi}) into Eq.\@ (\ref{eq:s1s2_cont}), we obtain the \enface complex signal: 
\begin{multline}
	s\rbra{\bmxp + \frac{\deltax}{2}}s^*\rbra{\bmxp - \frac{\deltax}{2}}
	= 
	\color{darkblue}
	\frac{w_0^4}{w^4} \exp \rbra{- \frac{\deltax^2}{w^2}}
	\iint D(\bmxpp) B^2(\bmxpp) \\ 
	\color{darkblue}
	\times \exp \abra{- \frac{4}{w^2}(\bmxp-\bmxpp)^2}
	\exp i \abra{- \frac{2 n k_0}{R} \deltax \cdot \rbra{\bmxp-\bmxpp}} 
        d \bmxpp
	\\  
	\quad + 
	\color{auburn}
	\frac{w_0^4}{w^4} 
	\exp \rbra{-\frac{\deltax^2}{w^2}}
	\iiiint D'(\bmxpp,\bmxqq) B(\bmxpp) B(\bmxqq) 
         \\
	\color{darkred}\times \exp\abra{- \frac{2}{w^2} (\bmxpp - \bmxqq) 
        \cdot\rbra{\frac{\bmxpp+\bmxqq}{2} -\deltax}}  
        e^{i \phi_{pq}}
        \\
        \color{darkred}\times \exp\abra{- \frac{4}{w^2} \rbra{\bmxp - \frac{\bmxpp+\bmxqq}{2}}^2} 
        \exp i\abra{\frac{2 n k_0}{R} 
        \rbra{\bmxpp - \bmxqq - \deltax }
        \cdot \rbra{\bmxp -\frac{\bmxpp + \bmxqq}{2}}}
	  \color{auburn} d \bmxpp d \bmxqq \\
	\color{auburn} -\mathrm{the\,\,\, first\,\,\, integral \,\,\, of \,\,\, R.H.S.}
	\label{eq:s1s2_Gauss}
\end{multline}
Similar to the conventional intensity-OCT image [Eq.\@ (\ref{eq:OCTInt_integral})], this \enface signal consists of the AI part [the first two lines of Eq.\@ (\ref{eq:s1s2_Gauss}), blue] and the CI part (the next four lines, brown and red).
The AI part forms an imaging formula in which the object to be measured is the same as the conventional intensity-OCT, that is, $D(\bmx)B^2(\bmx)$, the product of the scatterer density and the squared scattering potential. 
The PSF of this imaging formula is a complex PSF (the second line of the equation), which is the product of a Gaussian and a lateral linear phase slope. 
By contrast, the CI part is practically random, which is, again, similar to the conventional intensity-OCT.

Now we define the VDC image as the imaginary part of Eq.\@ (\ref{eq:s1s2_Gauss}) as
\begin{multline}
	\label{eq:VdicFull}
		\text{Im} \left[ \vdc \right]  
	= \color{darkblue} \frac{w_0^4}{w^4} \exp \rbra{- \frac{\deltax^2}{w^2}}
	\iint D(\bmxpp) B^2(\bmxpp) \\ 
	\color{darkblue} \times \exp \abra{- \frac{4}{w^2}(\bmxp - \bmxpp)^2}
	\sin \abra{- \frac{2 n k_0}{R} \deltax \cdot \rbra{\bmxp - \bmxpp}} d \bmxpp
	\\  
	\quad + \color{auburn} \frac{w_0^4}{w^4} 
	\exp \rbra{-\frac{\deltax^2}{w^2}}
	\iiiint D'(\bmxpp,\bmxqq) B(\bmxpp) B(\bmxqq)  \\
	\color{darkred} \times	\exp\abra{- \frac{2}{w^2} (\bmxpp-\bmxqq)\cdot \rbra{ \frac{\bmxpp+\bmxqq}{2} -\deltax}}  
	\exp\abra{- \frac{4}{w^2} \rbra{\bmxp - \frac{\bmxpp+\bmxqq}{2}}^2} \\
	\color{darkred}\times \sin \abra{\phi_{pq} + \frac{2 n k_0}{R} 
        \rbra{\bmxpp - \bmxqq - \deltax }
        \cdot \rbra{\bmxp - \frac{\bmxpp + \bmxqq}{2}}   
	}  \color{auburn} d \bmxpp d \bmxqq\\
	\color{auburn} -\mathrm{the\,\,\, first\,\,\, integral \,\,\, of \,\,\, R.H.S.}
\end{multline}
In this equation, the phase parts of Eq.\@ (\ref{eq:s1s2_Gauss}), i.e., the last exponential part of each integral, became sinusoidal function.
The last four lines of Eq.\@ (\ref{eq:VdicFull}) (brown and red) correspond to the CI part.
Similar to the case of the intensity-OCT formula, $\phi_{pq}$ and pseudo-PSF (red) are practically random; hence, CI becomes a random speckle pattern.
By omitting this random CI part, we can define the VDC image [$\mathrm{VDC}\rbra{\bmxp}$] and write it as the first integral of Eq.\@ (\ref{eq:VdicFull}): 
\begin{multline}
	\mathrm{VDC}\rbra{\bmxp}  
	\equiv \color{darkblue}\frac{w_0^4}{w^4} 
	\exp \rbra{- \frac{\deltax^2}{w^2}}\\
	\color{darkblue}
	\times \iint D(\bmx) B^2(\bmx)  
	\exp \abra{- \frac{4}{w^2}\rbra{\bmxp - \bmx}^2} 
	\sin \left[- \frac{2 n k_0}{R}\deltax \cdot \rbra{\bmxp - \bmx} \right]
	d \bmx \color{black}.
	\label{eq:vdc}
\end{multline}
We interpret this imaging formula as the convolution of the object $D(\bmx)B^2(\bmx)$ with an imaging PSF that is a Gaussian-windowed sine function.
This imaging PSF function is similar to the response function (i.e., the imaging PSF) of DIC \cite{Cogswell_1992} and also a differential operator used in digital image processing \cite{Canny:86}.
Hence, it works as a differential operator in VDC imaging and provides the differential contrast of the sample structure.

\begin{figure}
	\centering\includegraphics[width=9.9cm]{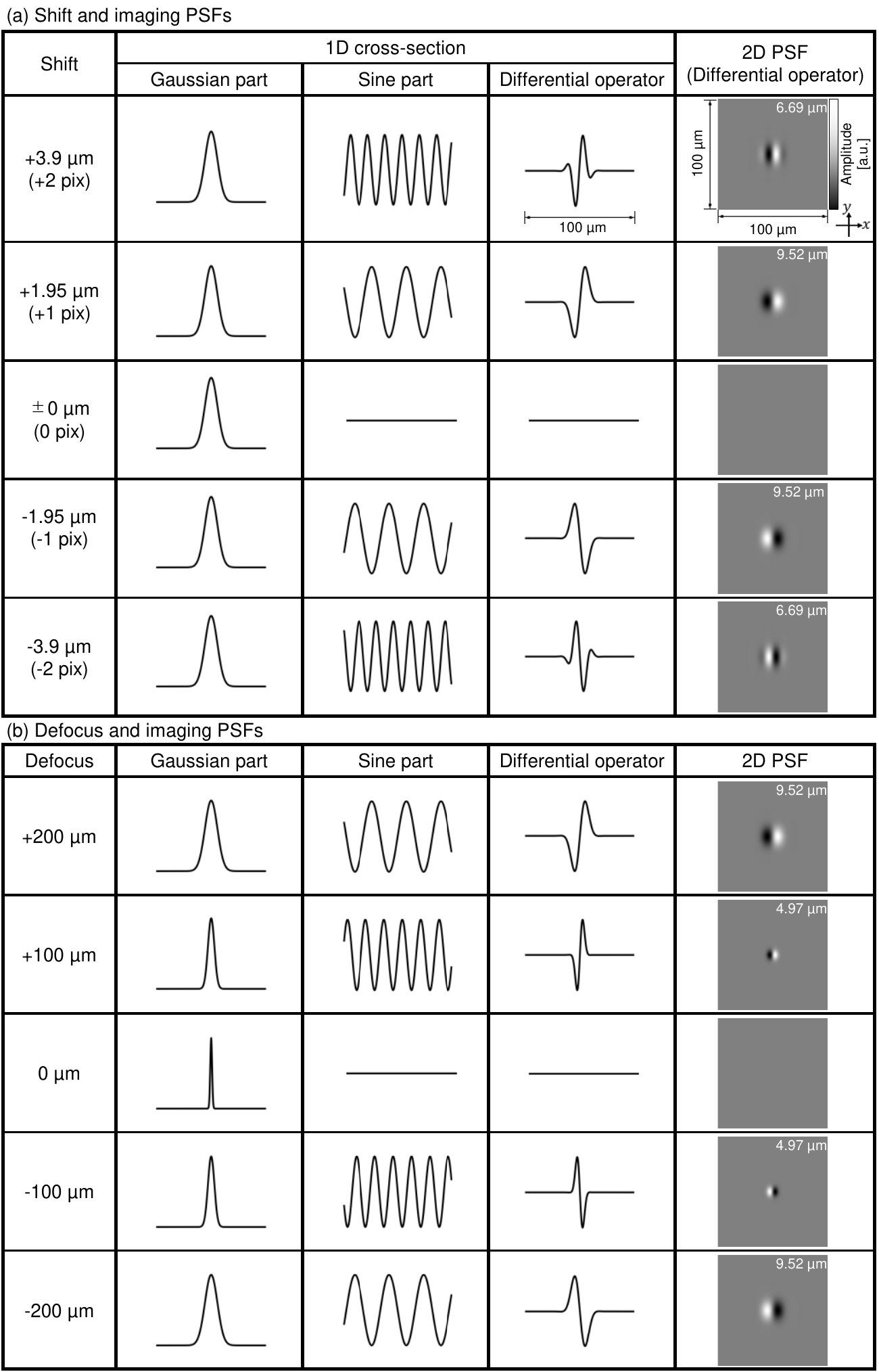}
	\caption{
        Examples of the imaging PSFs of VDC imaging with different shift amounts (a) and a different defocus amount (b).
        The one-dimensional plots are the profile of the two-dimensional \enface PSF at the center of the PSF in the shift direction.
        (a) The larger the shift amount, the higher the frequency of the sine part of the PSF and vice versa.
        When the shift is zero, the sine part becomes zero; hence, the image is not formed.
        The shift amount does not affect the Gaussian part.
        (b) The larger the defocus amount, the wider the Gaussian width and the higher the frequency of the sine part, and vice versa, provided that the defocus amount is larger than the Rayleigh length. 
		When the defocus amount is zero, the sine part becomes zero again; hence, the image is not formed.
		The appropriate selection of the shifting and defocus amounts leads to the PSF becoming a differential operator.
        The direction of differentiation can be flipped by flipping the shift and defocus directions.
        Note that the imaging PSF of VDC is analogous to the response function of DIC, and the shear amount and direction of DIC are analogous to the size and direction of the imaging PSF of VDC.
	}
	\label{fig:DerOperator}
\end{figure}
The shape of the imaging PSF of VDC (i.e., the differential operator) is altered by the defocus amount $z-z_0$ and shift $\deltax$.
Figure \ref{fig:DerOperator} exemplifies several PSFs with different shifts (a) and defocus amounts (b).
We computed PSFs using OCT system parameters identical to those in the experimental setup described in Section \ref{sec:VdcExperiment}: $w_0 = 2.41$ \um, $n= 1.38$, and $k_0=7.48 \times 10^6$ rad/m.
The shifting amounts were the integer multiples of the lateral pixel size (1.95 \um) and the defocus amounts were expressed as the distance from the focal depth, that is, $z-z_0$ in Eqs.\@ (\ref{eq:w}) and (\ref{eq:R}).
For Fig.\@ \ref{fig:DerOperator}(a), we set the defocus amount to 200 \um, whereas for Fig.\@ \ref{fig:DerOperator}(b), we set the shifting amount to 1.95 \um.
The one-dimensional plots are the profiles at the center of the PSF, i.e., $\bmxp = \bmx$, and are in the digital shifting direction.
The columns ``Gaussian part'' and ``sine part'' in the figure represent the Gaussian and sine parts within the integral of Eq.\@ (\ref{eq:vdc}), respectively.

As shown in Fig.\@ \ref{fig:DerOperator}(a), the shift affected the frequency of the sine function in the imaging PSF and did not affect the Gaussian function.
By contrast, the defocus amount affected both the Gaussian and sine functions of the imaging PSF [Fig.\@ \ref{fig:DerOperator}(b)].
Please note that the absolute defocus amount was larger than the Rayleigh length (21.7 \um) in the examples in the figure.

By properly selecting the shift and defocus amounts, we can obtain a good differential operator with an arbitrary differential distance, which is defined as the distance between two main positive and negative peaks.
The differential distances are shown at the top right of the 2D-PSF images.
(By contrast, an example of the ``inappropriate'' selection of these amounts is a shift that is too large, which fluctuates the operator within the Gaussian and an appropriate differential contrast cannot be obtained.)
When the defocus direction is reversed, that is, the plus/minus sign of the defocus is changed, the sine function and differential direction are flipped, as shown in Fig.\@ \ref{fig:DerOperator}(b).
When the shift is zero, the sine function becomes zero; hence, $\mathrm{VDC}(\bmxp) = 0$; 
that is, differential contrast is not obtained.
Similarly, when the defocus amount is zero, that is, $z=z_0$, $R$ becomes infinity and the sine part of the operator becomes $\sin 0 = 0$; hence, contrast is not obtained again.
Thus, not only the shift but also the defocus is mandatory for VDC.

In our particular implementation, we computationally apply defocusing as detailed in Section \ref{sec:compDefocus}; hence, both the shift and defocus amount can be selected after image acquisition.
It is also noteworthy that the size and direction of the differential operator are analogous to the shear amount and the shear direction of DIC imaging.
Because the shift and defocus amount can be set after image acquisition, VDC can generate multiple DIC-like images with arbitrary shear amounts and directions using post-acquisition signal processing.

\subsection{Experimental validation}
\label{sec:VdcExperiment}
\subsubsection{System, samples, and image acquisition protocol}
For the experimental validation of VDC imaging, we used a fiber-based spectral-domain OCT system\cite{Oida_2021}.
The light source was a superluminescent diode (SLD, Superlum, Ireland) with a center wavelength and bandwidth of 840 nm and 100 nm, respectively.
In the sample arm, the probe beam with a diameter of 4.0 mm passed through an objective (LSM02BB, Thorlabs, NJ) with a focal length of 18 mm and incident on the sample.
The axial resolution was 3.8 \um in tissue and the in-focus lateral resolution was 4.9 \um.
The spectral interference signal was acquired using a spectrometer (Cobra S800, Wasatch Photonics, NC).
The scanning speed was 50,000 lines/s with 2,048 spectral pixels. 

To reconstruct the OCT volume, spectral rescaling, numerical dispersion compensation, and fixed-pattern noise removal were applied.
The depth-independent bulk phase error caused by system imperfection and air turbulence was numerically corrected after the reconstruction of the OCT volume using the smart-integration-path method \cite{Oikawa_2020}. 
Note that, since our image processing exploits complex signals, high phase stability is required

\begin{figure}
	\centering\includegraphics{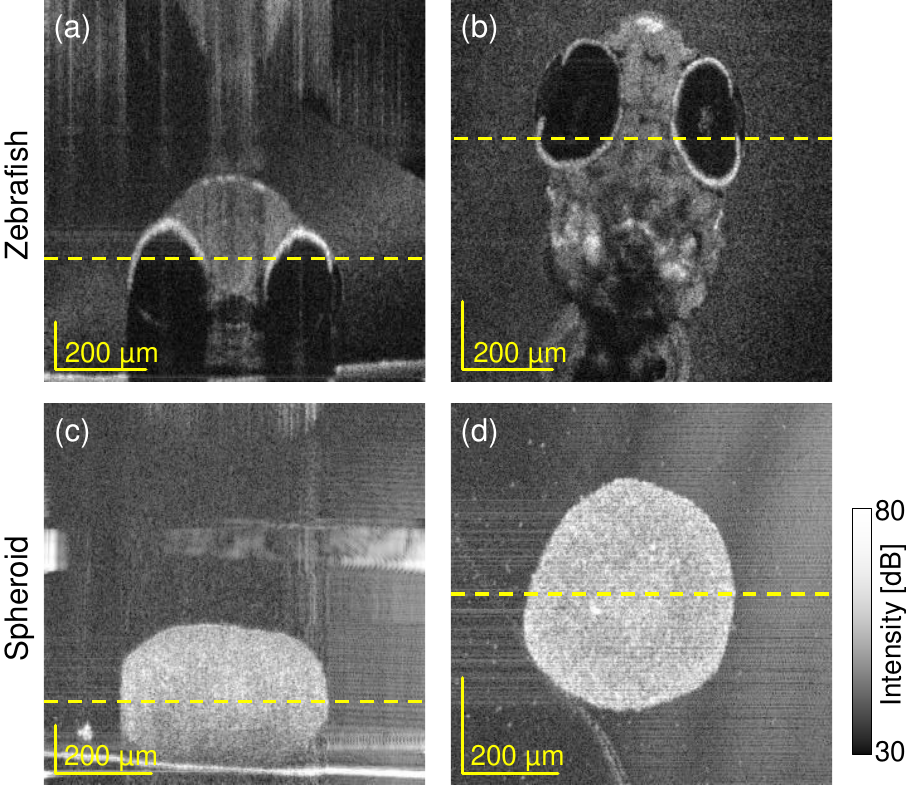}
	\caption{
		Cross-sectional (a, c) and \enface (b, d) intensity-OCT images of a zebrafish (a, b) and tumor spheroid (c, d).
		The yellow lines in the cross-sectional images indicate the depth positions of the \enface images.
		Similarly, the yellow lines in the \enface images indicate the positions of the cross-sectional images. 
	}
	\label{fig:IntenisityImage}
\end{figure}
An 8-day-old zebrafish larva and a human breast cancer (MCF7) spheroid were measured.
The zebrafish was anesthetized and embedded in agarose gel.
The lateral scanning areas were 1 mm $\times$ 1 mm for the zebrafish and 0.8 mm $\times$ 0.8 mm for the spheroid, with 512 $\times$ 512 A-lines.
Figure \ref{fig:IntenisityImage} shows examples of the cross-sectional and \enface OCT images of the (a) zebrafish and (b) spheroid.
The yellow dotted lines in each image indicate the depths of the \enface images and position of the cross-sectional images. 
These datasets were used for the validations in Section \ref{sec:results}.

\subsubsection{Processing of VDC images} 
\label{sec:processing}
\begin{figure}
	\centering\includegraphics[width=13cm]{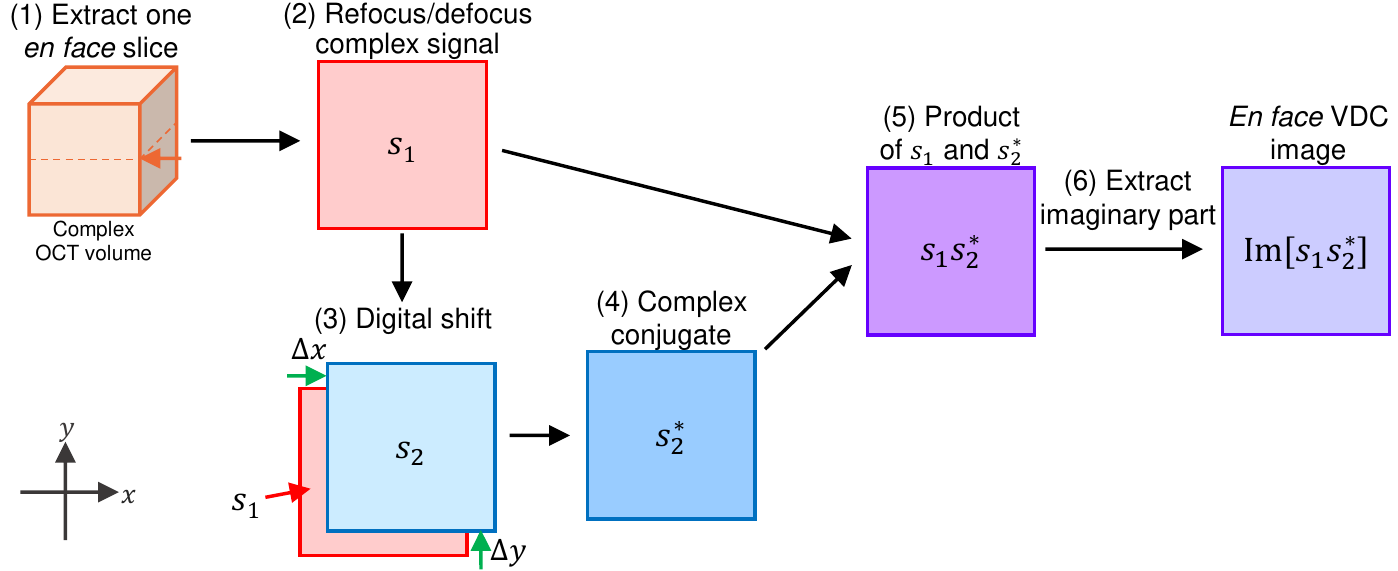}
	\caption{
		Processing flow of VDC imaging.
		(1) A complex \enface slice was extracted from a complex OCT volume.
		(2) Computational refocusing and defocusing were applied to obtain an arbitrarily defocused signal ($s_1$).
		(3) A digitally shifted copy of the \enface signal was generated ($s_2$) and (4) complex conjugated.
		(5) The product of $s_1$ and the complex conjugate of $s_2$ was computed. 
		(6) VDC was obtained as the imaginary part of the product ($s_1 s_{2}^*$).
	}
	\label{fig:ProcessingFlow}
\end{figure}
The VDC processing flow is summarized in Fig.\@ \ref{fig:ProcessingFlow}.
(1) An \enface plane was extracted from a complex OCT volume.
(2) The \enface signal was first refocused using computational refocusing based on a forward-propagation model \cite{Yasuno:06, Oikawa_2020} and then arbitrary numerical defocusing was applied (similar to that in Section 5.1.1 of Ref.\@ \cite{LidaZhu2022BOE}).
(3) This complex \enface OCT signal with arbitrary defocusing, $s_1(\bmxp)$, was copied and digitally shifted in the lateral direction to obtain $s_2(\bmxp)$.
(4, 5) Then the product of $s_1(\bmxp)$ and the complex conjugate of $s_2(\bmxp)$ was computed.
(6) Finally, the \enface VDC image was obtained as the imaginary part of the product.

\subsection{Results}
\label{sec:results}
\subsubsection{Shifting direction dependency of VDC imaging}
\begin{figure}
	\centering\includegraphics[width=13cm]{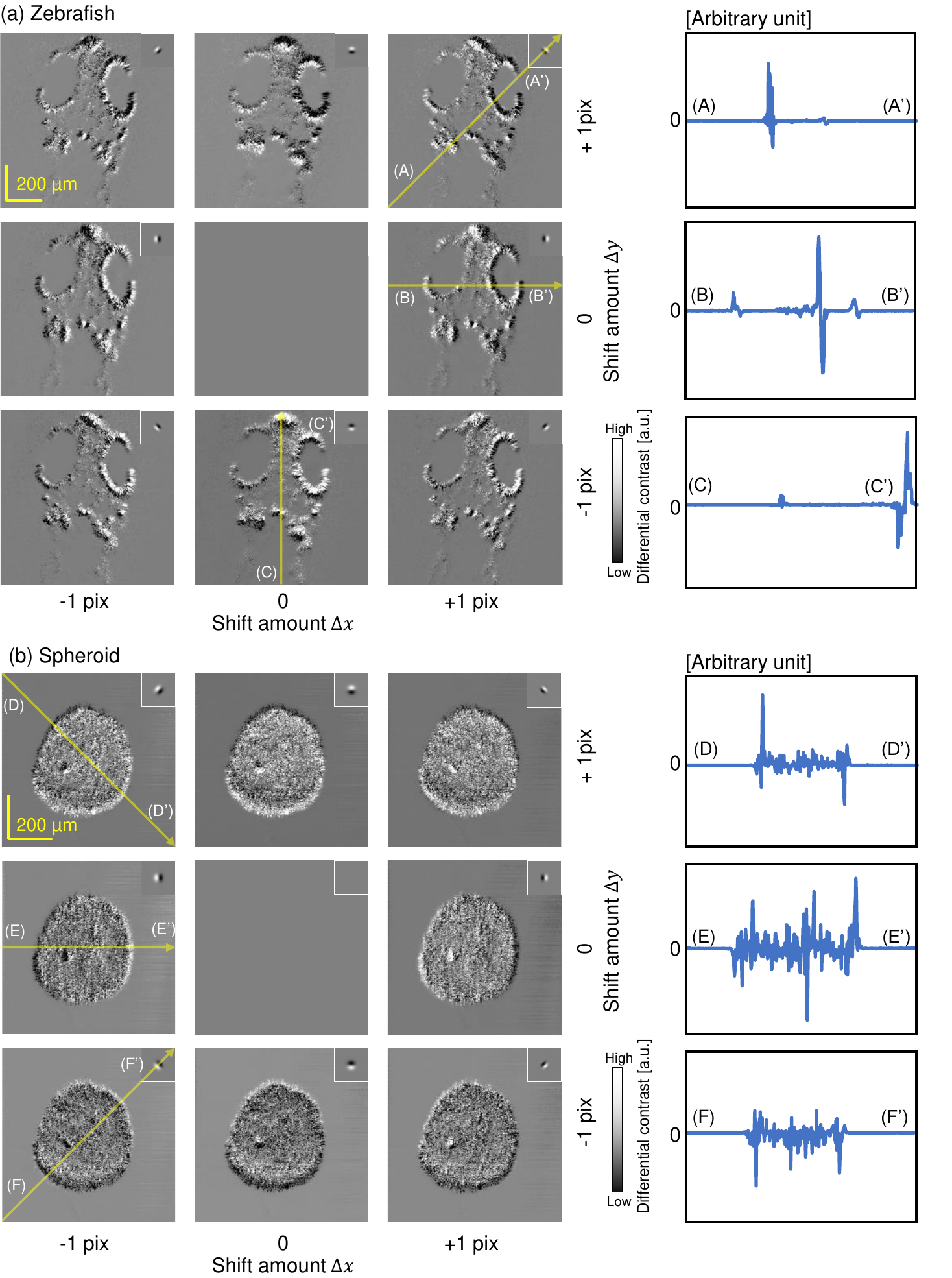}
	\caption{
		\Enface VDC images of the (a) zebrafish and (b) spheroid with multiple shift directions.
		The edge structures normal to the differential direction (i.e., shift direction) are emphasized.
		The insets show corresponding imaging PSFs.
		The plots on the right show some representative line profiles of the VDC images.
		The positions of the line profiles are indicated by long yellow arrows on the VDC images.
	}
	\label{fig:ShiftDir}
\end{figure}
The \Enface VDC images of the (a) zebrafish and (b) spheroid are summarized in Fig.\@ \ref{fig:ShiftDir}, which demonstrates the multi-directional differential imaging capability of VDC.
The plots on the right show some representative line profiles of the VDC images.
The positions of the line profiles are indicated by long yellow arrows on the VDC images.
The shifting amounts $\deltax = (\Delta x, \Delta y)$ were the combination of -1, 0, and +1 pixels (i.e., -1.95, 0.0, and +1.95 \um for the zebrafish and -1.56, 0.0, and +1.56 \um for the spheroid) for the $x$ and $y$ directions.
These shift amounts were selected to be the integer multiples of the pixel size to avoid sub-pixel interpolation of the complex OCT image.
The defocus amount was +200 \um for all images. 
Note that this defocus was not a physical defocus but a numerically induced defocus by using complex signal processing as mentioned in Section \ref{sec:processing}.
The combination of the shift and defocus amounts were selected to make the imaging PSF as a good differential operator. 
The small insets represent the imaging PSF.

As predicted by the imaging PSF shape of VDC, these images looked like the DIC image.
In the VDC images, the edge structures normal to the differential direction (shift direction) were particularly emphasized.
As expected, the VDC image was not obtained when there was no shift.
Note that, because DIC cannot be performed on a thick sample, such as the zebrafish and spheroid, a direct comparison of VIC and DIC cannot be performed.

\subsubsection{Shifting amount dependency of VDC imaging}
\begin{figure}
	\centering\includegraphics[width=13cm]{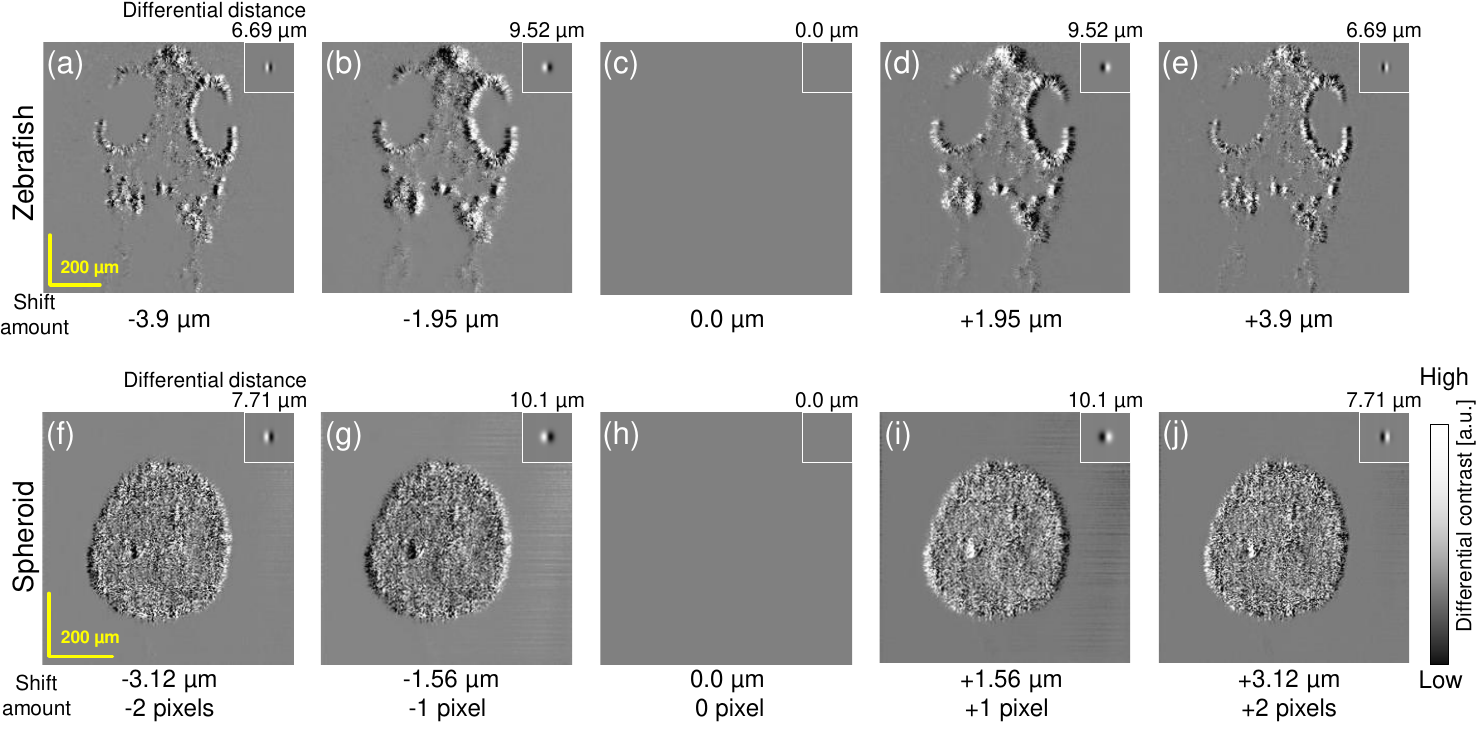}
	\caption{
		\Enface VDC images of the (a)--(e) zebrafish and (f)--(j) spheroid with shifting amounts of -2, -1, 0, +1, and +2 pixels.
        The small insets show the corresponding imaging PSFs and the values above them show the differential distances.
		$\pm$1-pix shift yielded  a larger differential distance than $\pm$2-pix shift, and hence emphasized larger structures than $\pm$2-pixel-shift images.
	}
	\label{fig:ShiftingAmt}
\end{figure}
By altering the shifting amount, VDC images with multiple differential distances are obtained as shown in Fig.\@ \ref{fig:ShiftingAmt}.
The shifting amounts were -2, -1, 0, +1, and +2 pixels (i.e., -3.9, -1.95, 0.0, +1.95, and +3.9 \um for the zebrafish and -3.12, -1.56, 0.0, +1.56, +3.12 \um for the spheroid) along the $x$-axis.
The defocus amount was +200 \um for all images.
The insets show the corresponding imaging PSFs and differential distances are shown above.

$\pm$1-pixel shifts [Fig.\@ \ref{fig:ShiftingAmt}(b), (d), (g), and (i)] yielded larger differential distances than $\pm$2-pixel shifts [Fig.\@ \ref{fig:ShiftingAmt}(a), (e), (f), and (j)].
Hence, $\pm$1-pixel-shift images emphasized larger structures than $\pm$2-pixel-shift images.
It is also noteworthy that the imaging PSFs of $\pm$2-pixel shifts exhibited small side peaks (ripples).
This may have deteriorated the VDC images.
When the shifting amount was zero, contrast was not obtained as expected [Fig.\@ \ref{fig:ShiftingAmt}(c) and (h)]. 

\subsubsection{Defocus amount dependency of VDC imaging}
\label{sec:compDefocus}
\begin{figure}
	\centering\includegraphics[width=13cm]{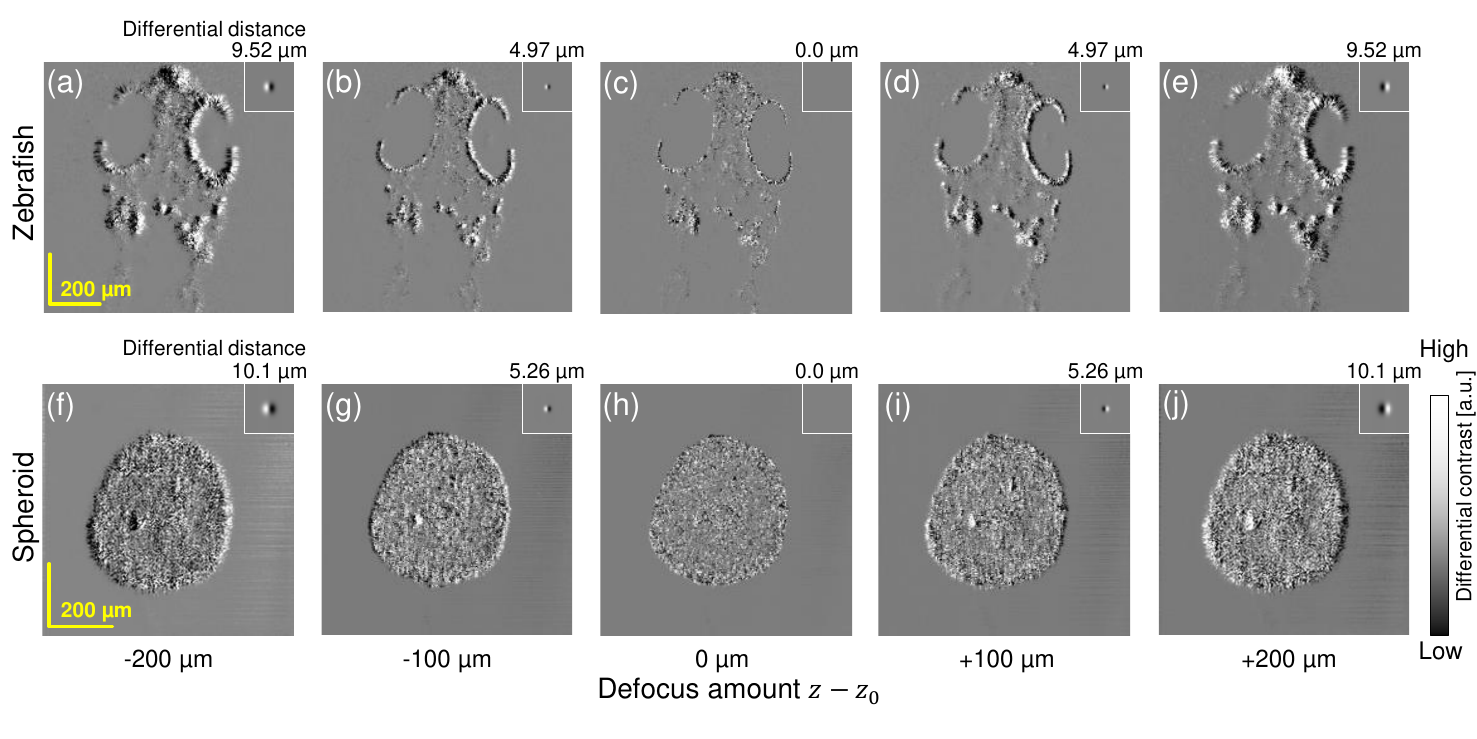}
	\caption{
		\Enface VDC images of the (a)--(e) zebrafish and (f)--(j) spheroid with defocus amounts of -200, -100, 0, +100, and +200 \um.
		The small insets show the corresponding imaging PSFs and the values above them show the differential distances.
		The differential distance increased as the amount of defocus increased.
		The differential direction reversed when the defocus direction reversed.
	}
	\label{fig:DefocusAmt}
\end{figure}
The effect of the defocus amount is demonstrated in Fig.\@ \ref{fig:DefocusAmt}, where the defocus amounts were -200, -100, 0, +100, and +200 \um.
The shifting amount was (+1, 0) pixel for the $(x,y)$-direction for all images.
The first and second rows show the VDC images of the zebrafish and spheroid, respectively.
The insets show the corresponding imaging PSFs and differential distances are shown above.

Increasing the absolute defocus amount increased the differential distance.
When the defocusing direction was reversed, the differential direction was also reversed. 
The VDC contrast became very low for the in-focus (0-\um defocus) and  [Fig.\@ \ref{fig:DefocusAmt}(c) and (e)].
It should be noted that, in theory, the VDC is expected to vanish at in-focus, but low contrast was visible in the images.
This residual contrast might be caused by the imperfection of numerical refocusing.

\subsubsection{Volumetric differential imaging}
\begin{figure}
	\centering\includegraphics{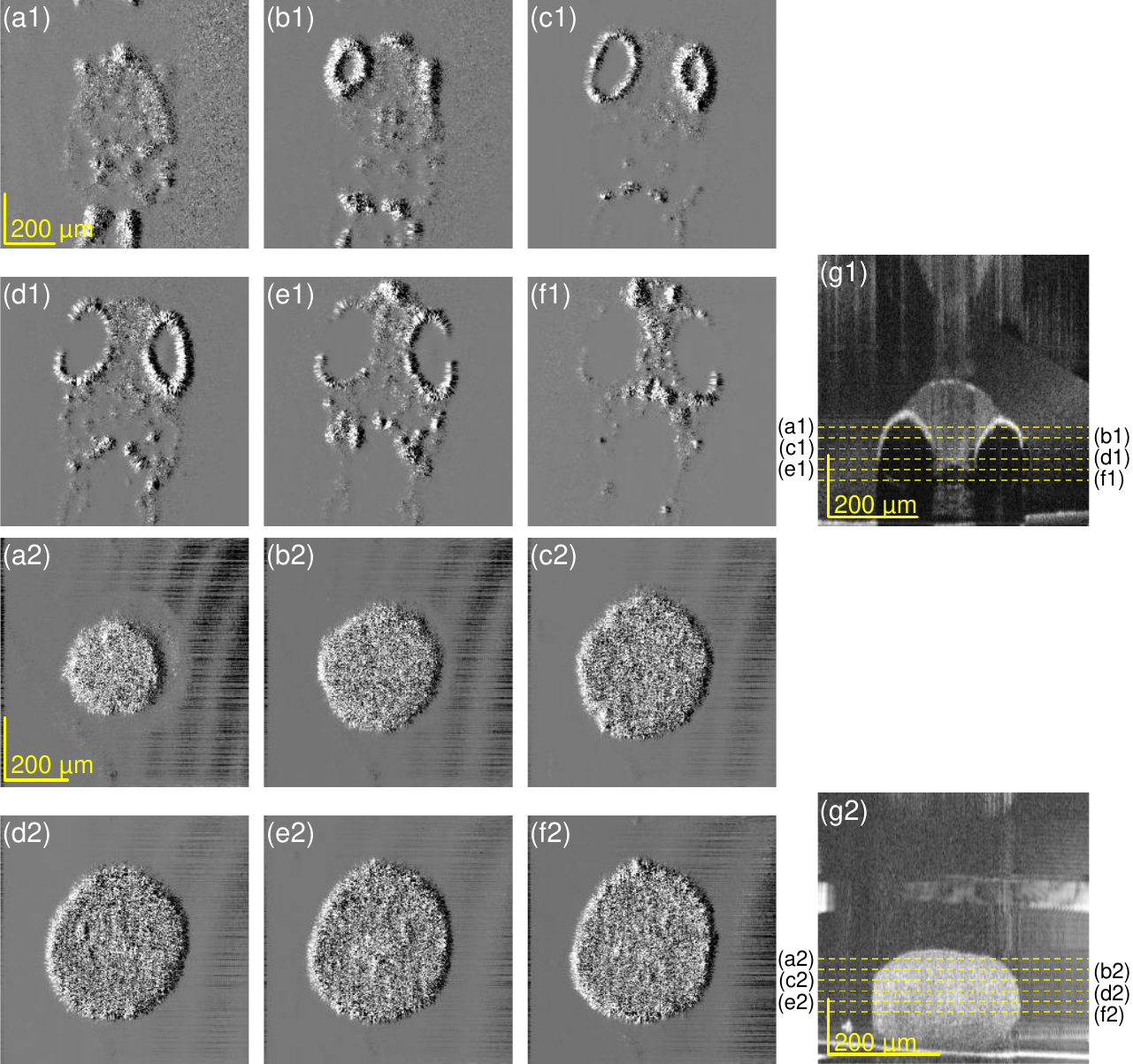}
	\caption{
		Volumetrically processed VDC images of the zebrafish and tumor spheroid.
		(a1--f1) \Enface VDC images at each depth of the zebrafish. 
		The depth positions are indicated in the cross-sectional OCT image (g1).
		(a2--g2) Same images for the tumor spheroid.
		The fly-through \enface images are available as Visualization-1 (zebrafish) and Visualization-2 (tumor spheroid).
	}
	\label{fig:volumeVDC}
\end{figure}

As shown above, VDC yielded DIC-like differential images.
However, unlike DIC, VDC achieved  volumetric differential imaging by processing each \enface plane in an OCT volume.
Figure \ref{fig:volumeVDC} shows examples of the zebrafish (a1--g1) and tumor spheroid (a2--g2).
(a1)--(f1) and (a2)--(f2) are the \enface VDC images extracted from the VDC volumes (from the upper side to the bottom side).
The depth locations of the \enface images are indicated in the OCT cross-sections (g1) and (g2). 
The defocus and shifting amounts were +200 \um and +1 pix along the $x$-axis for all images.
The fly-through \enface images for all depths are available as Visualization-1 (zebrafish) and Visualization-2 (tumor spheroid).

\section{Discussion}
\label{sec:discussion}
\subsection{Advantage of VDC imaging} 
As shown in Section \ref{sec:VDC}, VDC imaging provides images similar to DIC, but VDC imaging has two advantages over DIC.
First, VDC is volumetric and can image thick tissue.
Second, VDC can generate images with several differential conditions with a single acquisition;
that is, the differential direction and differential distance, which correspond to the shear direction and amount of DIC, respectively, can be controlled by the parameters of post-acquisition signal processing.
This means that VDC can be regarded as a volumetric DIC with post-hoc control of the shear direction and amount.

\subsection{Difference between VDC and DIC}
\label{sec:VDCvsDIC}
Despite their image similarity, the VDC and DIC have some fundamental differences in their contrast mechanisms.
DIC's contrast is mainly from the optical path length difference between two adjacent lateral points.
In addition, DIC is a transillumination microscope, and the sample is thin, low scattering, and typically, transparent.
And hence, the DIC's contrast is mainly based on the refractive index distribution of the sample.

On the other hand, VDC is obtained by OCT, which is a reflection (back-scattering) measurement modality, and the sample is assumed to exhibit high scattering.
And hence, VDC's contrast might be mainly based on the back-scattering.

One of the not-well-described effects in the current VDC theory is the effect of superior tissues.
VDC may be sensitive to the refractive index of the superior tissues, because it alters the phase of the complex \enface OCT signal at the depth of interest.
However, the present theory only considers an image at a particular depth, and does not formulate the cumulative effect of the superior tissues.
Extension of VDC theory to account for the cumulative effect is one of the future work.

In addition, three-dimensional imaging PSF of VDC has not yet been investigated.
As demonstrated by Villiger \etal \cite{Villiger2010JOSAA} and Ralston \etal \cite{Ralston2006JOSAA}, the three-dimensional imaging property of OCT can be formulated.
The extension of VDC imaging theory into three-dimensional space is also one of the future works.

Well-characterized-phantom based experimental validation might be helpful to further understand the imaging property of VDC.
However, for this validation, an inhomogeneous scattering phantom that has a slowly varying refractive index distribution and multiple domains with different scatterer densities are necessary.
The establishment of such characterized-phantoms and further experimental validation of VDC might be one of the future studies.

\subsection{Intensity based differential image generation} 
\label{sec:intensityDiff}
\begin{figure}
	\centering\includegraphics[width=13cm]{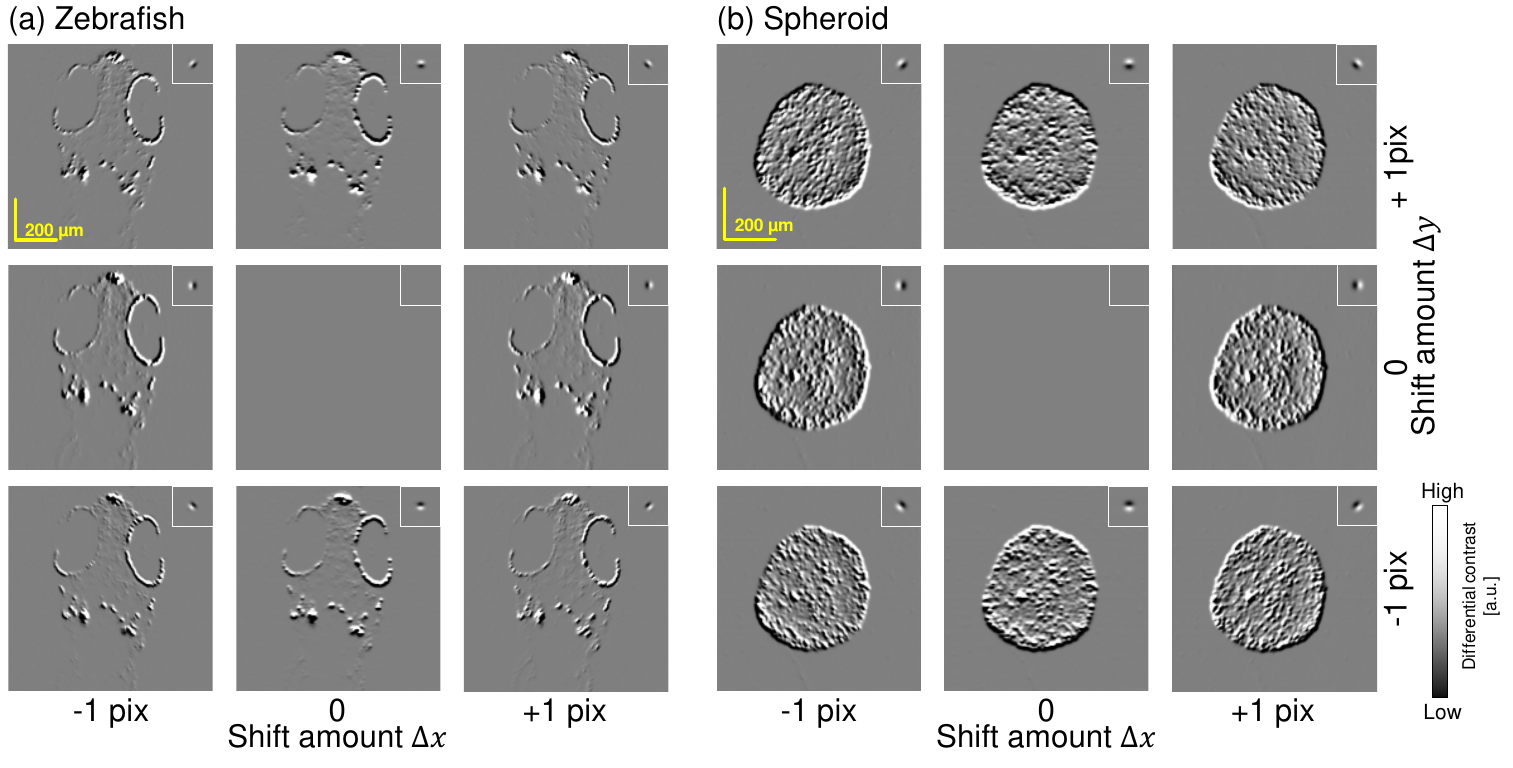}
	\caption{%
	Differential images generated from intensity based signal processing.
	The images were generated by convolving the linear-scale OCT intensity image with the imaging PSF of Fig.\@ \ref{fig:ShiftDir} which is shown also in the insets of this figure.
	The images corresponds to those of Fig.\@ \ref{fig:ShiftDir}.
	Since this operation differentiate the meaningful OCT image and speckle by the same differentiation operator, the signals originated from the meaningful OCT image and the speckle are hardly distinguished.
	}
	\label{fig:intensityDiff}
\end{figure}
It is noteworthy that, as far as we consider only the AI part, the differential contrast image can be obtained by one of the standard intensity-based image processing methods.
Figure \ref{fig:intensityDiff} shows the differential contrast images obtained by the intensity-based image processing corresponding to Fig.\@ \ref{fig:ShiftDir}.
Here the images were obtained by convolving the linear-scale OCT intensity image with the imaging PSF of Fig.\@ \ref{fig:ShiftDir}, and we denote these images as ``intensity differential images.''

Two differences are noted between VDC (Fig.\@ \ref{fig:ShiftDir}) and the intensity differential images (Fig.\@ \ref{fig:intensityDiff}).
First, the resolution of this intensity differential image might be slightly worse than that of VDC.
It is because the original intensity-OCT image used for the intensity differential image has already been affected by the intensity PSF [$P_a^2(\bmx)$ of Eq.\@ (\ref{eq:OCTInt_integral})].
However, the imaging PSF of VDC is much larger than the intensity PSF due to the numerical defocusing.
The defocus amount was 200 \um in this particular case, while the Rayleigh length is 21.7 \um.
And hence, this reduction of the resolution might be negligible.

The second and more significant difference is that the intensity differential images exhibit more granular appearance than the VDC images.
This difference can be explained by the effect of the CI part.
In the intensity differential images, CI part, i.e., speckle is also differentiated by the same differentiation kernel with the meaningful AI part.
And hence, ``differential speckle image'' is overlaid on the meaningful differential images.
Since both of them have the same resolution, it is hard to distinguish the differential speckle from the meaningful differential image.

On the other hand, the CI part of the VDC [i.e., the last two lines of Eq.\@ (\ref{eq:VdicFull})] does not have an evident PSF.
It has only a pseudo-PSF [see the paragraph just after Eq.\@ (\ref{eq:VdicFull})], and the width and the magnitude of pseudo-PSF alters for all combinations of the contributing scatterers.
And hence, the speckle-originated signal (CI part) appears quite differently from the meaningful differential image (AI part) in VDC.
Quantitative comparison of visibility of specific image features in specific applications between VDC intensity-based image-processing methods might be one of the future works in the transnational research.

\subsection{Artifacts in Doppler OCT}
\label{sec:Doppler}
Doppler OCT enables flow imaging in tissue\cite{Wang:07}.
Among several variations of Doppler OCT, conventional versions compute the phase difference between adjacent A-lines in a B-scan to obtain a Doppler OCT image \cite{White:03, Leitgeb2003OpEx_Doppler}.
This version of Doppler OCT is known to exhibit a phase noise pattern caused by the structure of the sample, that is, the structural phase artifact.
This structural phase artifact can be mathematically described by extending the formulation of VDC as follows:

This version of Doppler OCT can be written by modifying Eq.\@ (\ref{eq:s1s2_Gauss}) to $s\rbra{\bmxp + \deltax/2}$\xspace$s^*\rbra{\bmxp - \deltax/2}$\xspace$e^{i \varphi_D}$, where the newly added phase $\varphi_D$ is the Doppler phase shift and $\deltax$ is the separation of the adjacent A-lines.
The AI part of this modified equation becomes
\begin{multline}
	\mathrm{AI} =
	\frac{w_0^4}{w^4} \exp \rbra{- \frac{\deltax^2}{w^2}}
	\exp\rbra{i \varphi_D}
	\iint D(\bmxpp) B^2(\bmxpp) \\ 
	\times \exp \abra{- \frac{4}{w^2}(\bmxp-\bmxpp)^2}
	\exp i \abra{- \frac{2 n k_0}{R} \deltax \cdot \rbra{\bmxp-\bmxpp}} d \bmxpp
	\label{eq:Doppler}
\end{multline}
and the Doppler OCT signal is defined as the phase of this equation.

This formulation raises two points about the structural phase artifact.
First, if the static structures of the adjacent A-lines are identical, i.e., $s(\bmxp + \deltax/2) = s(\bmxp - \deltax/2)$, the Doppler OCT signal becomes the Doppler phase shift as 
\begin{equation}
\mathrm{angle}
\left[s\rbra{\bmxp + \frac{\deltax}{2}} s^*\rbra{\bmxp - \frac{\deltax}{2}} e^{\varphi_D}\right] = \varphi_D.
\end{equation}
Namely, no structural phase artifact appears if there is no structure.

Second, if there is no defocus, i.e., $R \rightarrow \infty$ in Eq.\@ (\ref{eq:Doppler}), the phase of Eq.\@ (\ref{eq:Doppler}) (the Doppler OCT signal) becomes the Doppler phase;
otherwise, the Doppler OCT signal is not identical to the Doppler phase.
It is noteworthy that, for the second point, we did not assume $s(\bmxp + \deltax/2) = s(\bmxp - \deltax/2)$,
i.e., the structural phase artifact does not appear, even if there is a structure as far as there is no defocus.
To summarize, the structural phase artifact appears only when the structure and defocus coexist.

Another structure-related issue is known for Doppler OCT; that is,
``structural decorrelation,'' which means that the measurement accuracy of the Doppler phase degrades as the distance between the adjacent A-lines ($\deltax$) increases.
This effect can be related to the first exponential part of Eq.\@ (\ref{eq:Doppler});
that is, the AI signal decreases as $\deltax$ increases and leads to a low signal-to-noise ratio, and hence, low phase measurement accuracy.

\subsection{Defocusing effect of computational multi-directional (CMD)-OCT}
\label{sec:CMD}
Oida \etal demonstrated point-scanning computational multi-directional (CMD)-OCT for qualitatively visualizing the microscopic phase structure of the sample \cite{Oida_2021}.
This method exploits the phase angle of $\vdcinline$, and hence is similar to our VDC method.

Although it was empirically known that the contrast of CMD-OCT becomes very low if the image is in focus \cite{Oida_bios}, its theoretical basis was not known.
This low contrast has now been explained by Eq.\@ (\ref{eq:s1s2_Gauss}) of the VDC formulation;
that is, if there is no defocus (i.e., $R \rightarrow \infty$), the phase of the AI part (meaningful part) of $\vdcinline$ becomes zero.
This fact explains the contrast vanishing of CMD-OCT at in-focus.

\subsection{The relation of our and previous OCT imaging theories}
Zhou \etal has formulated the OCT signal as the collection of contributions from discrete scatterers (Section 7 of \cite{KCZhou2021AOP}), and gave a similar representation with our discrete equation of intensity-OCT [Eq.\@ (\ref{eq:OctIntensityGrouped})]. 
Namely, they represented the OCT image and speckle as two independent additive components.

Our work added two points to their work.
At first, we derived a continuous form of the intensity-OCT imaging equation [Eq.\@ (\ref{eq:OCTInt_integral})].
This equation clarified that the meaningful OCT image is the convolution of the object [$D(\bm{x})B^2(\bm{x})$] and a PSF of OCT intensity image $P^2_a(\bm{x})$.
Second, our work accounts for the phase of complex PSF.
It has enabled the design of VDC (Section \ref{sec:VDC}), interpretations of artifacts in Doppler OCT (Section \ref{sec:Doppler}), and understand the imaging property of computational CMD-OCT (Section \ref{sec:CMD}).

Our OCT formulation does not represent the three-dimensional imaging property, similar to that discussed in Section \ref{sec:VDCvsDIC} for VDC.
It is because we have assumed that the lateral and axial imaging properties are independent of each other.
Namely, low numerical aperture (NA) was tacitly assumed.
Villiger \etal \cite{Villiger2010JOSAA} and Ralston \etal \cite{Ralston2006JOSAA} have formulated the three-dimensional imaging property of OCT, in which the lateral and axial imaging properties interact with each other.
The extension of our OCT formulation to account for the axial imaging property and its interaction with the lateral imaging property might be a future work.

Another point to be discussed is the applicability of our formulation strategy to full-field (FF-) OCT.
There are several works of formulation of FF-OCT imaging.
For example, Tricoli \etal formulated imaging property of partially coherent FF-OCT \cite{Tricoli2019JOSAA}, and Barolle \etal gave a theoretical model which recapitulates the manifestation of aberrations in FF-OCT \cite{Barolle2021OpEx}.
Our formulation is potentially adoptable to FF-OCT.
However, after Eq.\@ (\ref{eq:PSF}), it is assumed that the illumination and collection pupil functions are identical.
And this assumption is valid for point-scanning OCT but not true for FF-OCT with spatially incoherent light sources.
We can make our theory applicable to FF-OCT by reformulating it from Eq.\@ (\ref{eq:OCTInt_integral}).

\section{Conclusion}
We presented a new mathematical model called DSM to represent a sample to be measured by OCT.
Using this model, we reformulated lateral OCT imaging and also designed a new imaging method called VDC imaging.

Using the OCT reformulation, the meaningful signal and speckle were successfully formulated as two different mathematical entities.
We also found that the meaningful signal and speckle originated from the auto- and cross-numerical interactions of the scatters' signal contribution, respectively.

VDC imaging is a combination of physical OCT imaging and subsequent complex signal processing, and we derived an imaging formula that expresses the physical and numerical process simultaneously.
The imaging PSF that appeared in this formula suggested that VDC imaging yields an image similar to that yielded by DIC.
We validated the VDC method by measuring biological samples.

Finally, the formulation strategy has the newly explained mathematical mechanism of the well-known but previously not well-explained artifacts of Doppler OCT, and the imaging property of CMD-OCT.

The presented formulation strategy may be applicable to other OCT-based imaging methods in the future.
Although the present formulation focuses on the lateral imaging property, a similar approach based on DSM can be used to extend this formulation for axial and three-dimensional imaging properties.

\section*{Funding}
Core Research for Evolutional Science and Technology (JPMJCR2105);
Japan Society for the Promotion of Science (21H01836, 22K04962, 22F22355, 22H05556, 21H02997);
Japan Science and Technology Agency (JPMJMI18G8);
Austrian Science Fund (Schr\"odinger grand J4460).

\section*{Disclosures}
Tomita, Makita, Morishita, Abd El-Sadek, Mukherjee, Lichtenegger, Yasuno: Nikon (F), Sky Technology (F), Yokogawa Electric Corp. (F), Kao Corp. (F), Topcon (F).
Fukutake: Nikon (E).
Tamaoki, Bian, Kobayashi: None.
Mori, Matsusaka: None.

\section*{Data availability}
The data that support the findings of this study are available from the corresponding author upon reasonable request.

\appendix
\section*{Appendix}
In this appendix, we derive Eq.\@ (\ref{eq:s1s2_cont}) in Section \ref{sec:VDC}.
The derivation is based on a logical flow similar to that used to obtain Eq.\@ (\ref{eq:OCTInt_integral}) from Eq.\@ (\ref{eq:OCTInt_sum}).
First, we calculate the product of an \enface complex OCT signal and its digitally laterally shifted complex conjugate as
\begin{align}
        & \vdc \nonumber \\
        & \qquad= \left[ \sum_{i=1}^{N} B(\bmxi)     e^{i\phi_i} 
        \aPsf{\bmxp -\bmxi + \frac{\deltax}{2}}\, e^{i \varphi \rbra{\bmxp -\bmxi + \frac{\deltax}{2}} }\right]\nonumber \\
         &\qquad \quad\times \left[ \sum_{j=1}^{N} B(\bmxj) e^{-i\phi_j} 
         \aPsf{\bmxp -\bmxj - \frac{\deltax}{2}}\, e^{-i \varphi \rbra{\bmxp -\bmxj - \frac{\deltax}{2}} }\right] 
         \nonumber \\
        &\qquad = \color{darkblue}\sum_{i=1}^{N} B^2(\bmxi) 
        \aPsf{\bmxp - \bmxi+\frac{\deltax}{2}} 
        \aPsf{\bmxp - \bmxj-\frac{\deltax}{2}}
        e^{i \left[\varphi \rbra{\bmxp -\bmxi + \frac{\deltax}{2}}
        - \varphi \rbra{\bmxp -\bmxi - \frac{\deltax}{2}} \right]}
        \nonumber \\  
	&\qquad \quad+ \color{auburn} \underset{(i \neq j)}{\sum_{i=1}^{N}\sum_{j=1}^{N}}  
	B(\bmxi) B(\bmxj) e^{i(\phi_i-\phi_j)}
        \nonumber \\
        &\qquad \qquad \color{auburn}\times
        \color{darkred}\aPsf{\bmxp -\bmxi + \frac{\deltax}{2}} 
	\aPsf{\bmxp -\bmxj - \frac{\deltax}{2}}  
        e^{i \left[\varphi \rbra{\bmxp -\bmxi + \frac{\deltax}{2}}
        - \varphi \rbra{\bmxp -\bmxj - \frac{\deltax}{2}} \right]}\color{black}.
        \label{eq:VDC_discrete}
\end{align}	
The first summation of the final expression (the third last line of the equation, blue) is for the case of $i=j$, i.e., for the AI group.
The second summation (the last two lines of the equation, brown and red) is for the case of $i \neq j$, that is, the CI group.
Using the Dirac delta function, $\bmxpp$, and $\bmxqq$, the physical spatial coordinates in the sample, Eq.\@ (\ref{eq:VDC_discrete}), are rewritten as  
\begin{multline}
	\label{eq:VDC_dirac}
       \vdc \\
	= \color{darkblue} \iint \sum_{i=1}^{N} \delta \rbra{\bmxpp - \bmxi}
        B^2(\bmxpp) 
        \aPsf{\bmxp - \bmxpp + \frac{\deltax}{2}}
        \aPsf{\bmxp - \bmxpp - \frac{\deltax}{2}}
        \\
        \color{darkblue} \times \exp {i \left[\varphi \left(\bmxp - \bmxpp+\frac{\deltax}{2} \right)    - \varphi \left(\bmxp - \bmxpp-\frac{\deltax}{2} \right) \right]}
        d\bmxpp \\  
         + \color{auburn} \iiiint  \sum_{i=1}^{N} \sum_{j=1}^{N} 
         \delta \rbra{\bmxpp - \bmxi, \bmxqq - \bmxj}
         B(\bmxpp) B(\bmxqq) \color{darkred} e^{i(\phi_{pq})} \\
	\color{darkred} \times \aPsf{\bmxp - \bmxpp+\frac{\deltax}{2}}
	\aPsf{\bmxp - \bmxqq - \frac{\deltax}{2}}
	e^{i \left[ \varphi 
        \left(\bmxp - \bmxpp+\frac{\deltax}{2} \right)
		-  \varphi \left(\bmxp - \bmxqq-\frac{\deltax}{2} \right) \right]}
	d\bmxpp d\bmxqq \\
	\color{auburn} -\mathrm{the\,\,\, first\,\,\, integral \,\,\, of \,\,\, R.H.S.}
\end{multline}
The subtraction on the last line is because of the same reason as that in Eq.\@ (\ref{eq:OctInterContinuous}).
By substituting the scatterer density maps [Eqs.\@ (\ref{eq:D}) and (\ref{eq:D'})] into this equation, we obtain the continuous form, i.e., Eq.\@ (\ref{eq:s1s2_cont}).

\bibliography{reference}

\end{document}